\documentclass[12pt]{iopart}
\usepackage{iopams}
\usepackage{graphicx}
\usepackage[T1]{fontenc}
\usepackage{ae}
\usepackage{color}
\usepackage[latin1]{inputenc}

\newcommand{\di}{\unitlength0.5cm\begin{picture}(1,1)
\put(0.,0.2){\circle*{0.2}}
\put(0.5,0.2){\circle*{0.2}}
\put(0,0.2){\line(1,0){0.5}}
\end{picture}}

\newcommand{\dii}{\unitlength0.6cm\begin{picture}(1,1)
\put(0.,0.2){\circle*{0.2}}
\put(0.5,0.2){\circle*{0.2}}
\put(0,0.1){\line(1,0){0.5}}
\put(0,0.3){\line(1,0){0.5}}
\end{picture}}

\newcommand{\diii}{\unitlength0.6cm\begin{picture}(1,1.5)
\put(0.,0.2){\circle*{0.2}}
\put(0.5,0.2){\circle*{0.2}}
\put(0,0.2){\line(1,0){0.5}}
\put(0.25,0.2){\oval(0.5,0.3)}
\end{picture}}

\newcommand{\ti}{\unitlength0.6cm\begin{picture}(1,1.5)
\put(0.,0.0){\circle*{0.2}}
\put(0.5,0.0){\circle*{0.2}}
\put(0.25,0.65){\circle*{0.2}}
\put(0,0.0){\line(1,0){0.5}}
\put(0,0.0){\line(1,3){0.23}}
\put(0.5,0.0){\line(-1,3){0.23}}
\end{picture}}

\newcommand{\tii}{\unitlength0.6cm\begin{picture}(1,1)
\put(0.,0.2){\circle*{0.2}}
\put(0.5,0.2){\circle*{0.2}}
\put(1,0.2){\circle*{0.2}}
\put(0,0.1){\line(1,0){0.5}}
\put(0,0.3){\line(1,0){0.5}}
\put(0.5,0.2){\line(1,0){0.5}}
\end{picture}}

\newcommand{\tiiii}{\unitlength0.6cm\begin{picture}(1,1.5)
\put(0.,0.0){\circle*{0.2}}
\put(0.5,0.0){\circle*{0.2}}
\put(0.25,0.55){\circle*{0.2}}
\put(0,-0.1){\line(1,0){0.5}}
\put(0,0.1){\line(1,0){0.5}}
\put(0,-0.1){\line(1,3){0.24}}
\put(0.5,-0.1){\line(-1,3){0.24}}
\end{picture}}

\begin{document}


\title[Moments of Heisenberg Hamiltonians ]
{Moments of general Heisenberg
Hamiltonians up to sixth order }

\author{Heinz-J\"urgen Schmidt$^1$
\footnote[3]{Correspondence should be addressed to
hschmidt@uos.de}
, Andre Lohmann$^2$ and Johannes Richter$^2$ }

\address{$^1$ Universit\"at Osnabr\"uck, Fachbereich Physik,
Barbarastr. 7, D - 49069 Osnabr\"uck, Germany\\
$^2$Institut f\"ur Theoretische Physik, Otto-von-Guericke-Universit\"at Magdeburg,\\
PF 4120, D - 39016 Magdeburg, Germany }

\maketitle


\begin{abstract}
We explicitly calculate the moments $t_n$ of general Heisenberg
Hamiltonians up to sixth order. They have
the form of  finite sums of products of two factors, the first
factor being represented by a multigraph and the second factor being
a polynomial in the variable $s(s+1)$, where $s$ denotes the
individual spin quantum number. As an application we determine the
corresponding coefficients of the expansion of the free
energy and the zero field susceptibility in powers of the inverse temperature.
These coefficients can be written in a form which makes explicit their extensive character.
\end{abstract}

\section{Introduction}\label{sec:I}
The moments of the general Heisenberg Hamiltonian describing a
large class of spin systems represent valuable information for
various purposes. They can be used to check numerical exact or
approximate calculations of the energy spectrum of such systems.
Further they directly appear in the high temperature expansion
(HTE) of the partition function. Hence they can be used to
calculate the leading terms of the HTE of physical relevant
functions such as specific heat or magnetic susceptibility, see
e.~g.~\cite{DG74} or \cite{OHZ06}. Analytic expression for the moments in the
general Heisenberg case (arbitrary number of spins $N$ and
coupling constants $J_{\mu\nu}$, arbitrary spin quantum number $s$) are
interesting since they can be used to investigate the dependence
of physical properties on the coupling constants even in cases
where a numerical investigation is not possible due to large $N$
or $s$. It seems that such analytical expressions for moments have
only be published up to order three, see \cite{SSL01},
although
the polynomials which appear in such expressions are known up
to $8$th order, see \cite{DG74}. Those
papers which consider higher order expansions are usually confined
to special
coupling geometries or and/or $s=1/2$, see e.~g.~\cite{TWB10}-
\nocite{CGSM08} \nocite{FHW04} \nocite{OZ04a} \nocite{OZ04b}
\nocite{LKMW03}\nocite{ST02}\nocite{HL01}
\nocite{BEU00}\nocite{ZHO99}\nocite{ES98a}
\nocite{ES98b}\nocite{OB96}
\cite{KBJ96}.

In this article we only give the
analytical results for the moments $t_4,\,t_5,\, t_6$ without
explaining the method by which we obtained them. This will be done
in another paper. From these results we have determined the
coefficients of the susceptibility's HTE $c_4,\,c_5,\,c_6$ by a
method which is briefly sketched.
Similarly, the coefficients $a_n$ of the free energy's
series expansion (or, more precisely, of $-\beta F(\beta)$) are calculated up to $6$th order in $\beta$.
We have checked our results by
comparison with numerical results for spin systems with random
coupling coefficients and $N\le 7$ and $s=1/2,1$. Moreover, our
expression for $c_4$ has been checked by comparison with the
published results
for chains \cite{Jetal00} and $J_1-J_2$ square lattices \cite{Retal03}\cite{MBP03}.\\

\newpage

\section{Definitions}\label{sec:D}
We consider the following notations:\\

\begin{tabular}{ll}
$N$ & Number of spins\\
${\mathcal S}_N$ & Group of permutations $\pi:\{1,2,\ldots,N\}\rightarrow \{1,2,\ldots,N\}$\\
$s=\frac{1}{2},1,\frac{3}{2},\ldots$ & Single spin quantum number\\
$r\equiv s(s+1)$& Abbreviation\\
$J_{\mu\nu}=J_{\nu\mu},\,1\le\mu\neq\nu\le N$ & Coupling constants\\
$\mathbf{s}_\mu$ & Spin vector operator of the $\mu$th spin\\
$\mathbf{S}=\sum_{\mu}\bf{s}_\mu$ & Total spin vector\\
$\mathbf{S}^{(i)},\;\;i=1,2,3$ & $i$th component of the total spin vector\\
$H=\sum_{\mu<\nu}J_{\mu\nu}\bf{s}_\mu\cdot\bf{s}_\nu$ & Heisenberg Hamilton operator\\
$(2s+1)^N$ & Dimension of the  Hilbert space\\
$t_n=\frac{\mbox{\scriptsize Tr}(H^n)}{(2s+1)^N}$& Normalized moments of $H$\\
$\mu_n=\frac{\mbox{\scriptsize Tr}(\mathbf{S}^{(3)2}H^n)}{(2s+1)^N}$& Normalized magnetic moments of $H$\\
$\beta=\frac{1}{k\,T}$ & Inverse temperature\\
$\chi(\beta)=\beta \frac{\mbox{\scriptsize Tr}(\mathbf{S}^{(3)2}\exp(-\beta H))}
{\mbox{\scriptsize Tr}(\exp(-\beta H))}$& Zero field susceptibility\\
$\chi(\beta)=\sum_{n=1}^{\infty} c_n \beta^n$& High temperature expansion of $\chi(\beta)$\\
$-\beta F(\beta)=\ln\left( \mbox{Tr } e^{-\beta H}\right)$ & Free energy $F(\beta)$\\
$-\beta F(\beta)=\sum_{n=0}^{\infty} a_n \beta^n$& High temperature expansion of $F(\beta)$
\end{tabular}
\vspace{1cm}

\begin{table}
\caption{Multigraphs ${\mathcal G}_\nu$ \label{table1}}
\begin{center}
\begin{tabular}{||l|l||l|l||l|l||l|l||}
\hline \\
{$\nu$ }& ${\mathcal G}_\nu$ &$\ldots$ & $\ldots$ &$\ldots$ & $\ldots$ &$\ldots$ & $\ldots$\\
\hline
\hline\\
 {$1$}  &{\includegraphics[width=2cm]{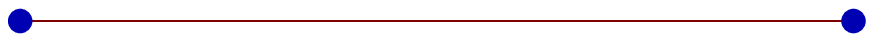}} &$9$  &\includegraphics[width=2cm]{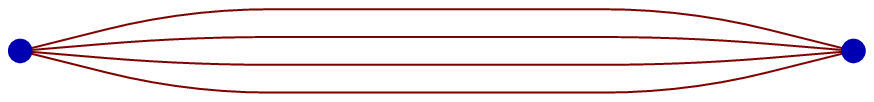}
 &$17$  &\includegraphics[width=2cm]{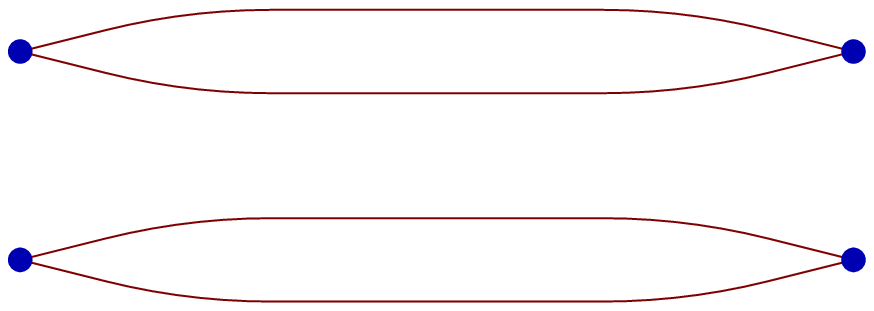} &$25$  &\includegraphics[width=2cm]{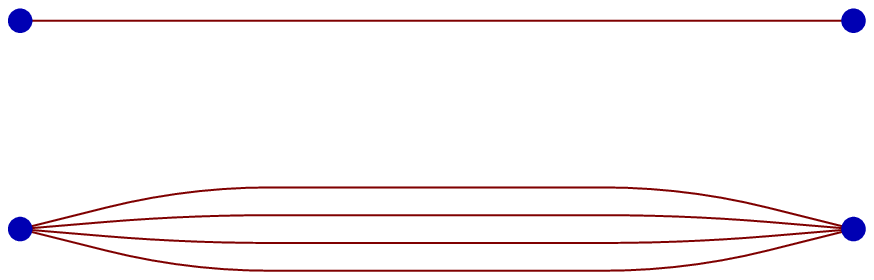}\\
\hline\\
$2$  &\includegraphics[width=2cm]{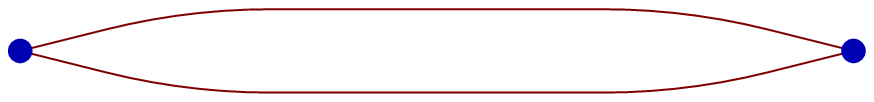} &$10$
&\includegraphics[width=2cm]{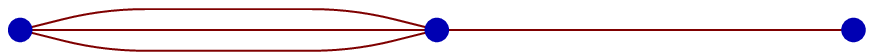}
 &$18$  &\includegraphics[width=2cm]{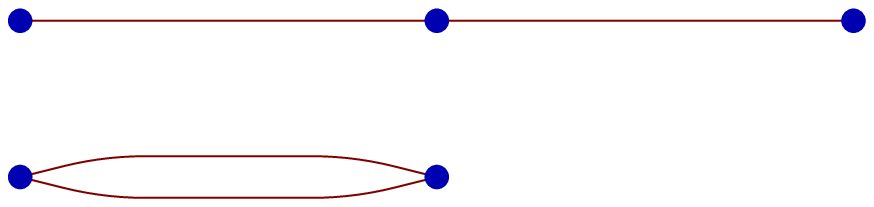} &$26$  &\includegraphics[width=2cm]{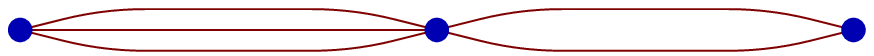} \\
\hline\\
$3$  &\includegraphics[width=2cm]{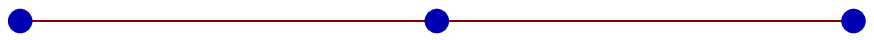} &$11$
&\includegraphics[width=2cm]{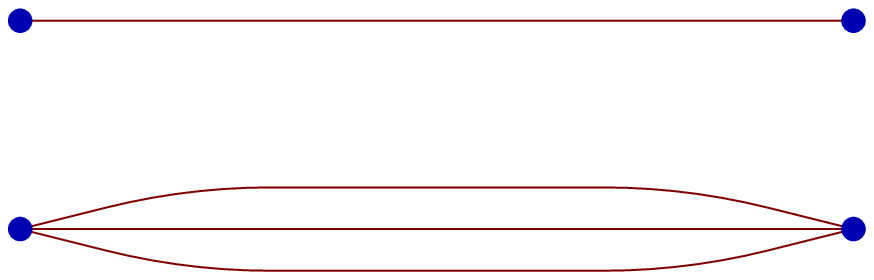}
 &$19$  &\includegraphics[width=2cm]{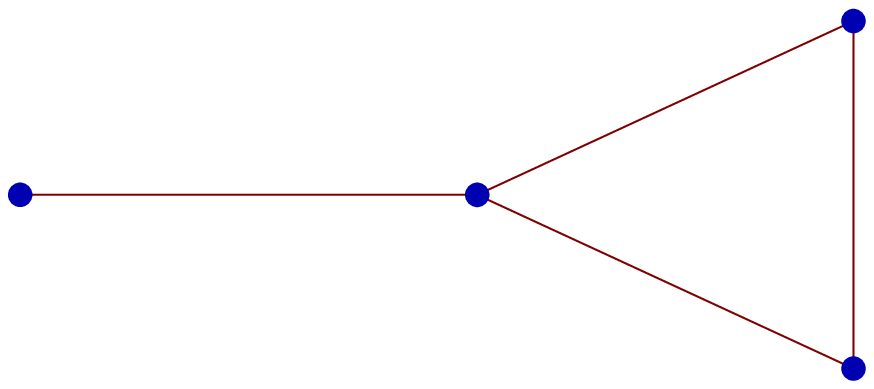} &$27$  &\includegraphics[width=2cm]{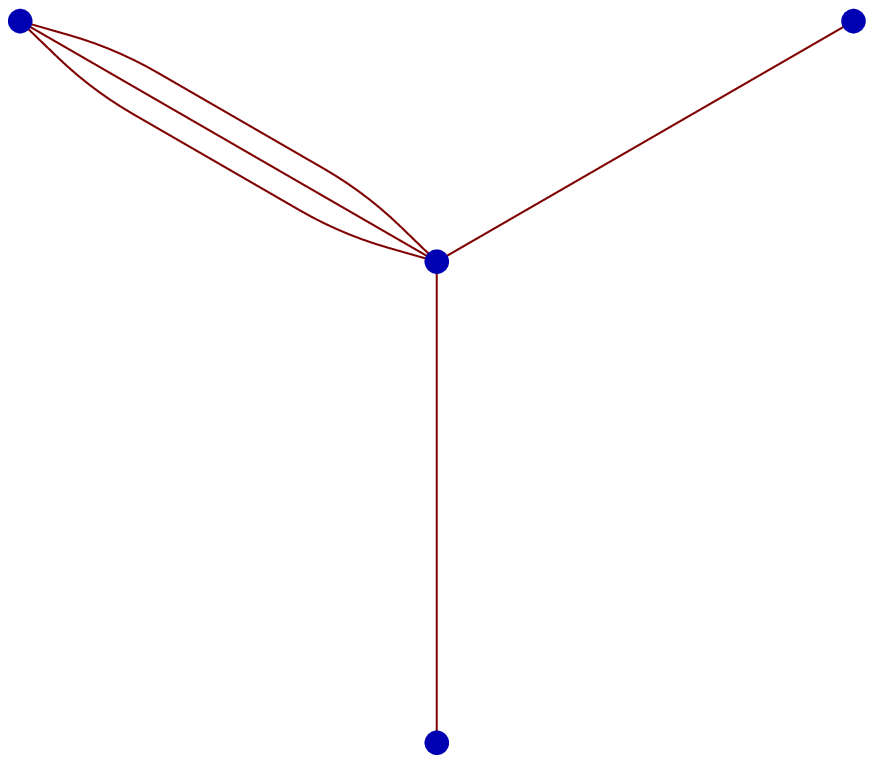}\\
\hline\\
$4$  &\includegraphics[width=2cm]{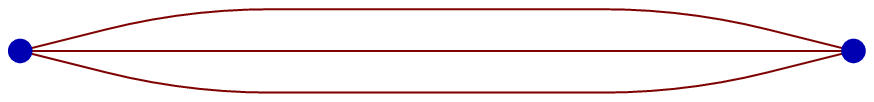}&$12$
&\includegraphics[width=2cm]{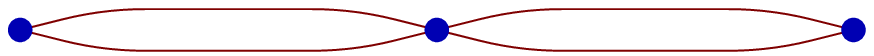}
 &$20$  &\includegraphics[width=2cm]{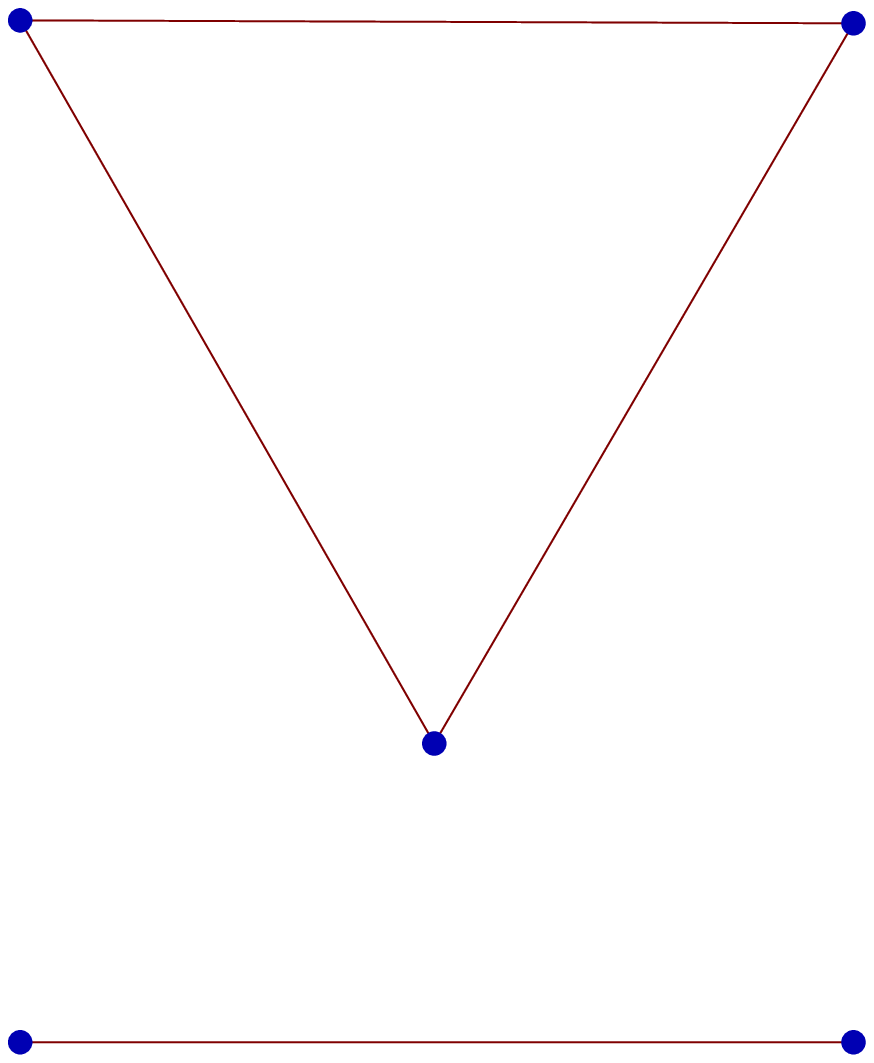} &$28$  &\includegraphics[width=2cm]{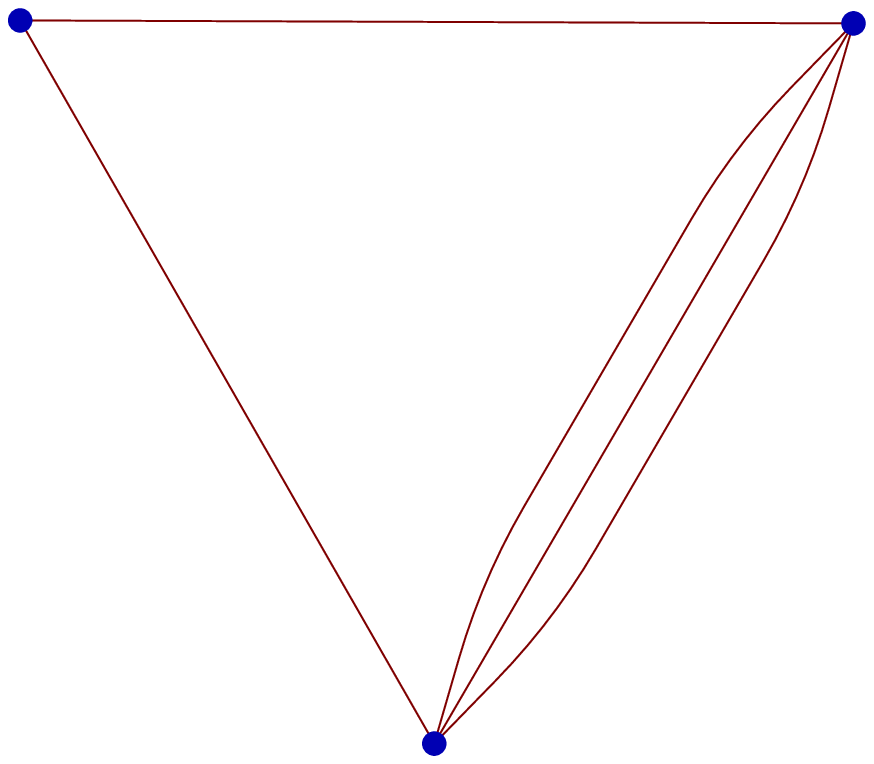}  \\
\hline\\
$5$  &\includegraphics[width=2cm]{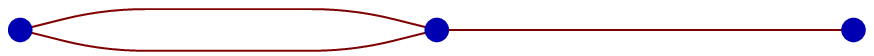}&$13$
&\includegraphics[width=2cm]{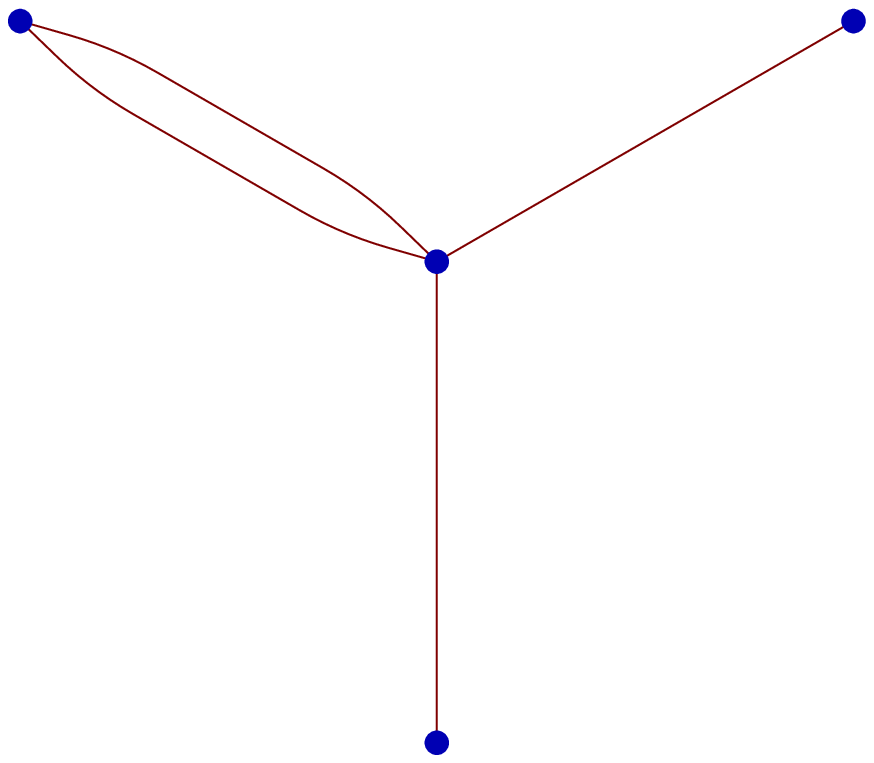}
 &$21$  &\includegraphics[width=2cm]{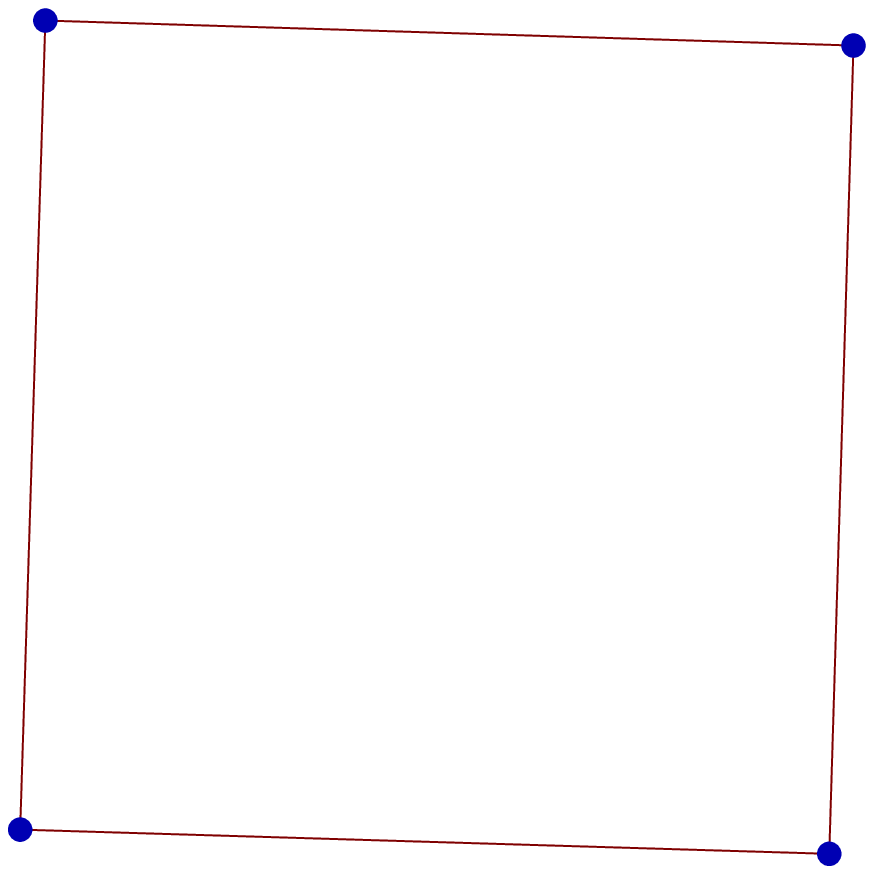} &$29$  &\includegraphics[width=2cm]{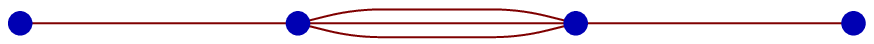} \\
\hline\\
$6$  &\includegraphics[width=2cm]{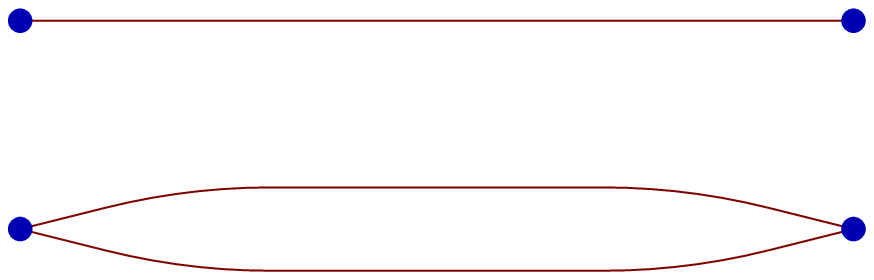}&$14$
&\includegraphics[width=2cm]{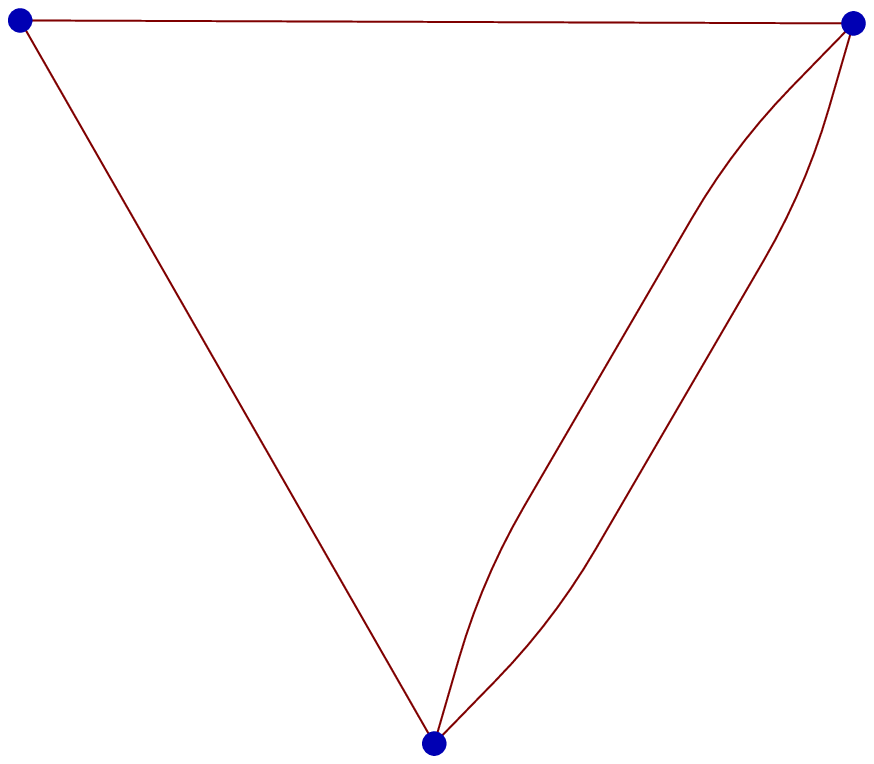}
 &$22$  &\includegraphics[width=2cm]{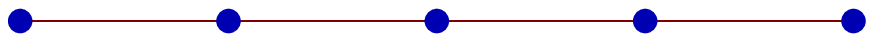} &$30$  &\includegraphics[width=2cm]{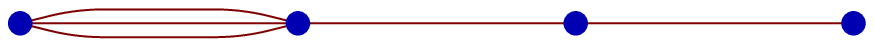}\\
\hline\\
$7$  &\includegraphics[width=2cm]{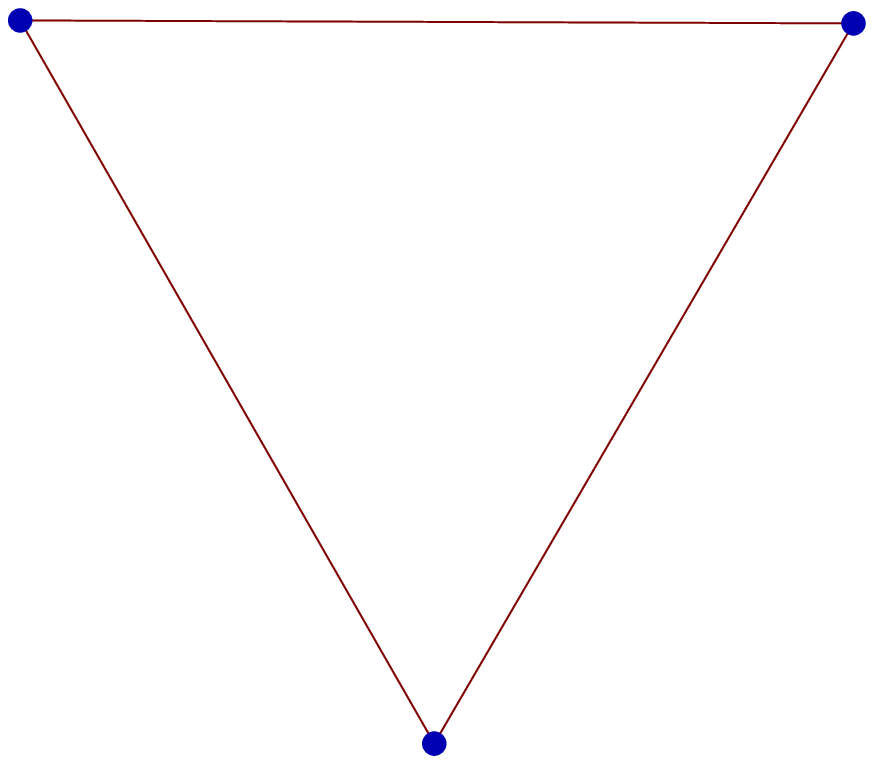}&$15$
&\includegraphics[width=2cm]{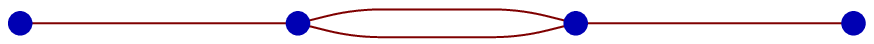}
 &$23$  &\includegraphics[width=2cm]{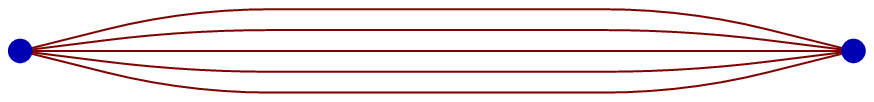} &$31$  &\includegraphics[width=2cm]{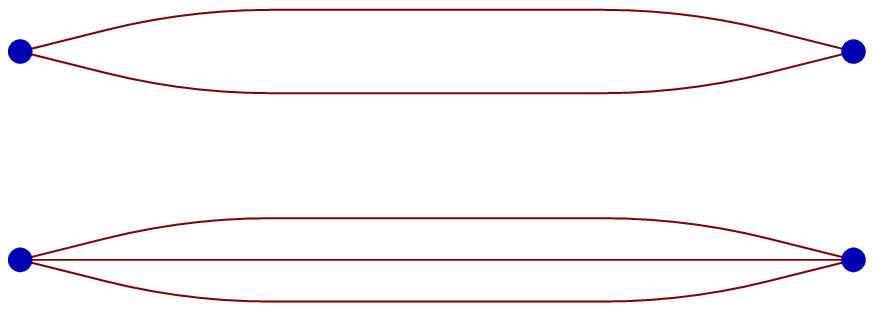}  \\
\hline\\
$8$  &\includegraphics[width=2cm]{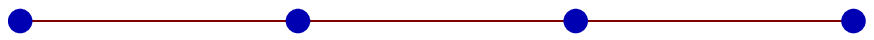} &$16$
&\includegraphics[width=2cm]{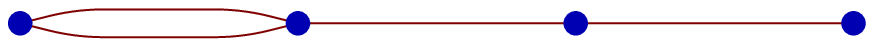}
 &$24$  &\includegraphics[width=2cm]{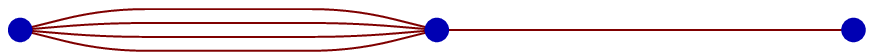} &$32$  &\includegraphics[width=2cm]{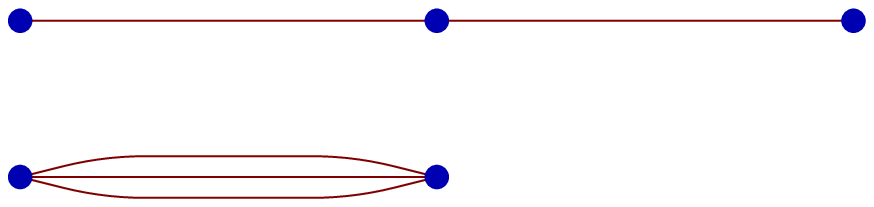} \\
\hline
\end{tabular}
\end{center}
\end{table}

\begin{table}
\begin{center}
\begin{tabular}{||l|l||l|l||l|l||l|l||}
\hline \\
$\nu$ & ${\mathcal G}_\nu$ &$\ldots$ & $\ldots$ &$\ldots$ & $\ldots$ &$\ldots$ & $\ldots$\\
\hline
\hline\\
 $33$  &\includegraphics[width=2cm]{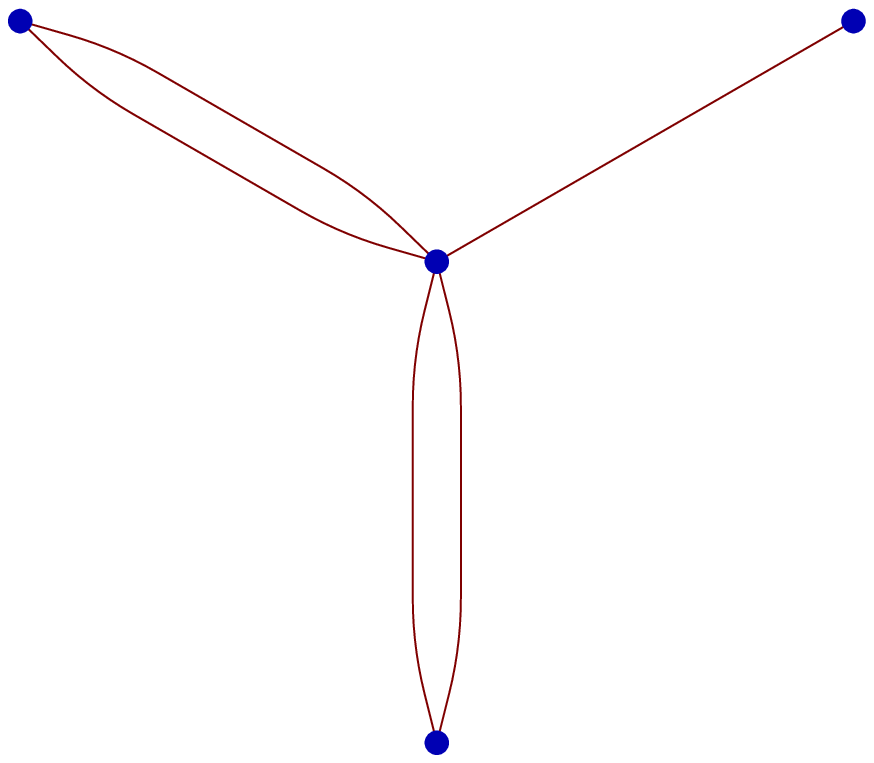} &$41$  &\includegraphics[width=2cm]{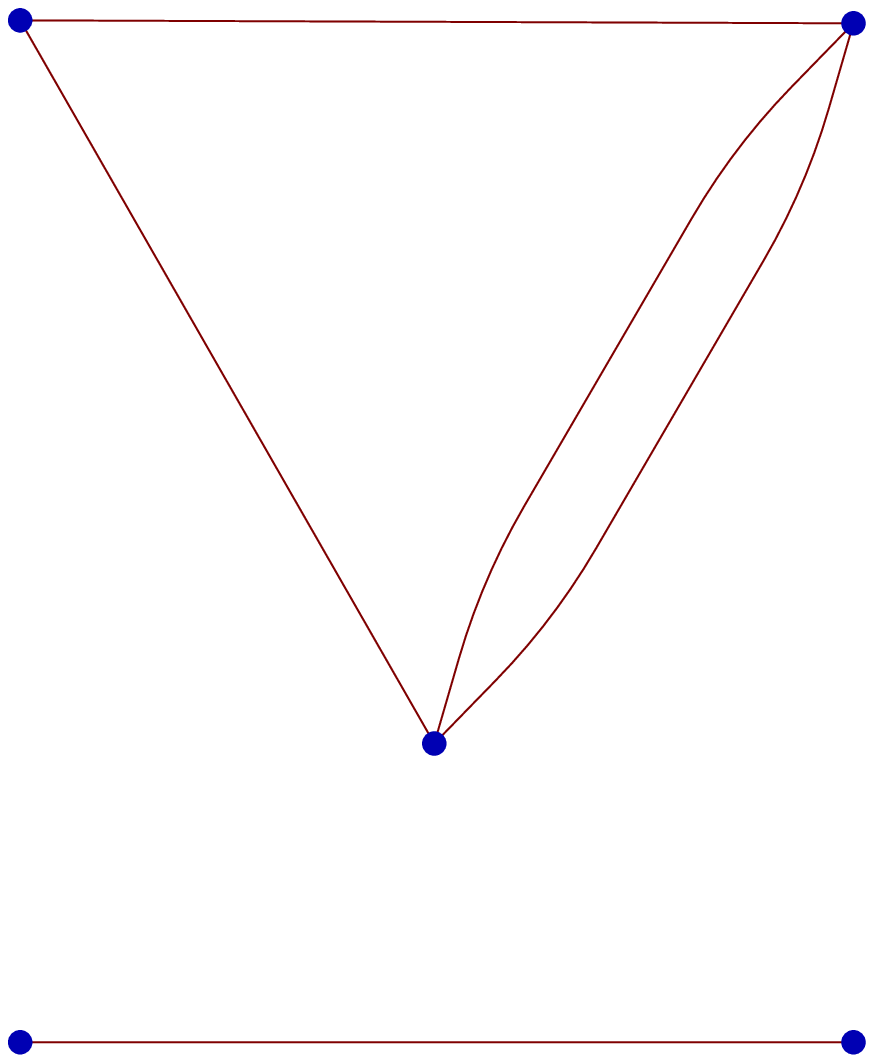}
 &$49$  &\includegraphics[width=2cm]{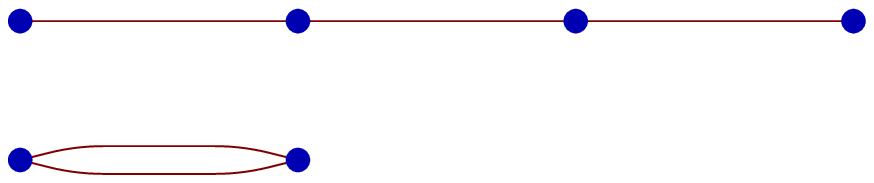} &$57$  &\includegraphics[width=2cm]{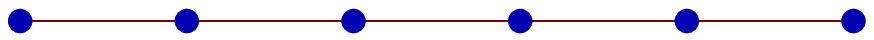}\\
\hline\\
$34$  &\includegraphics[width=2cm]{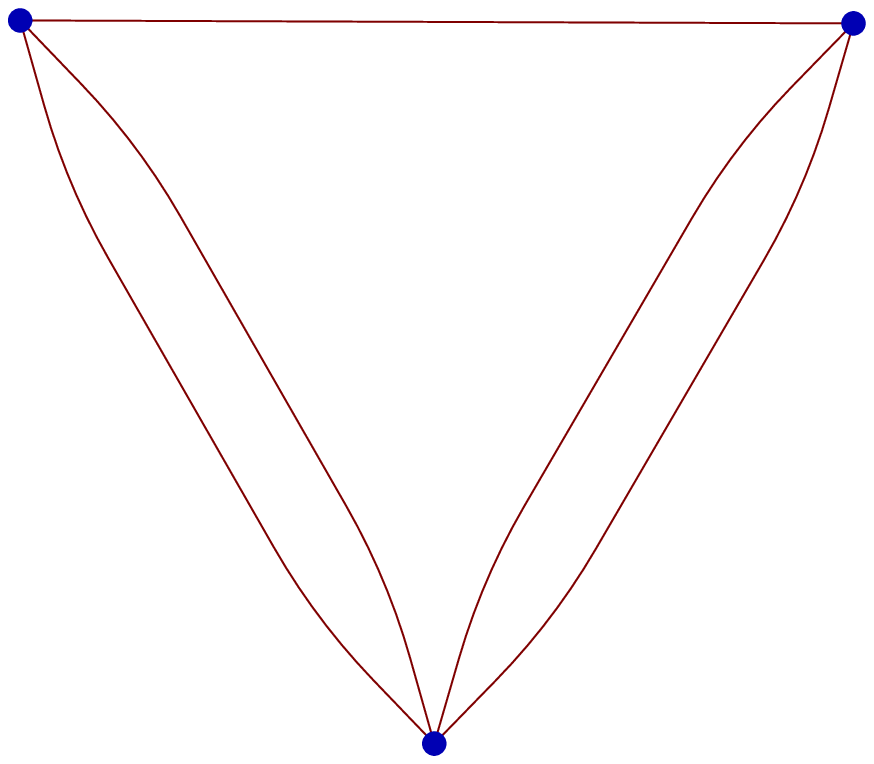} &$42$
&\includegraphics[width=2cm]{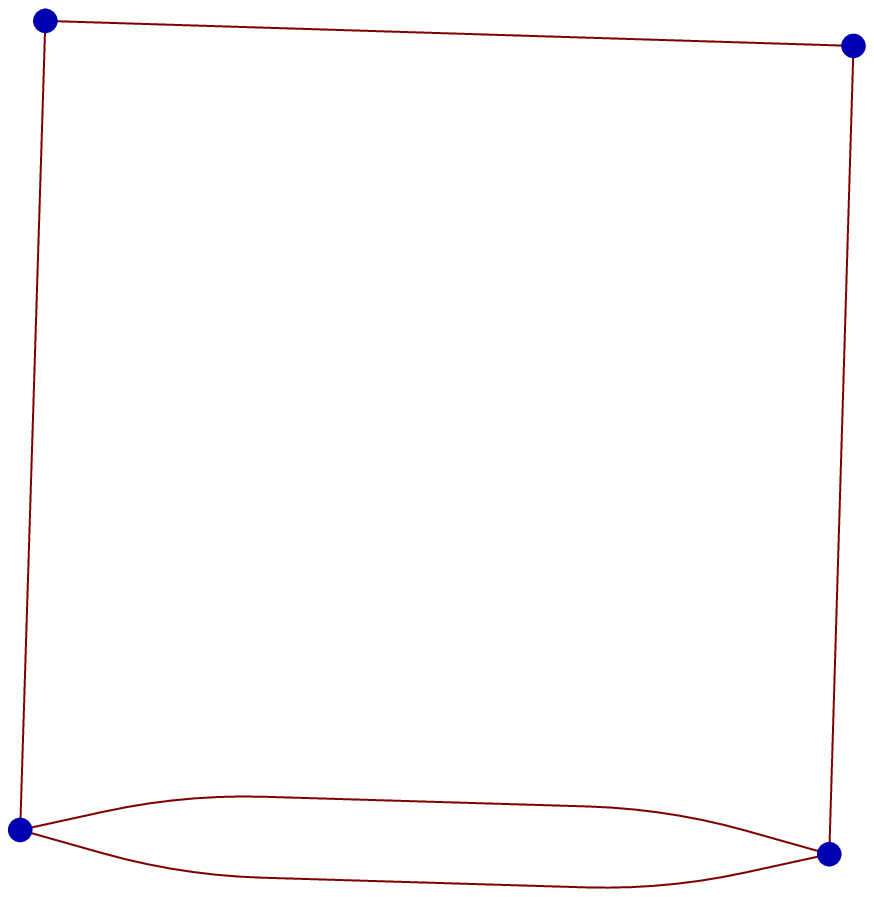}
 &$50$  &\includegraphics[width=2cm]{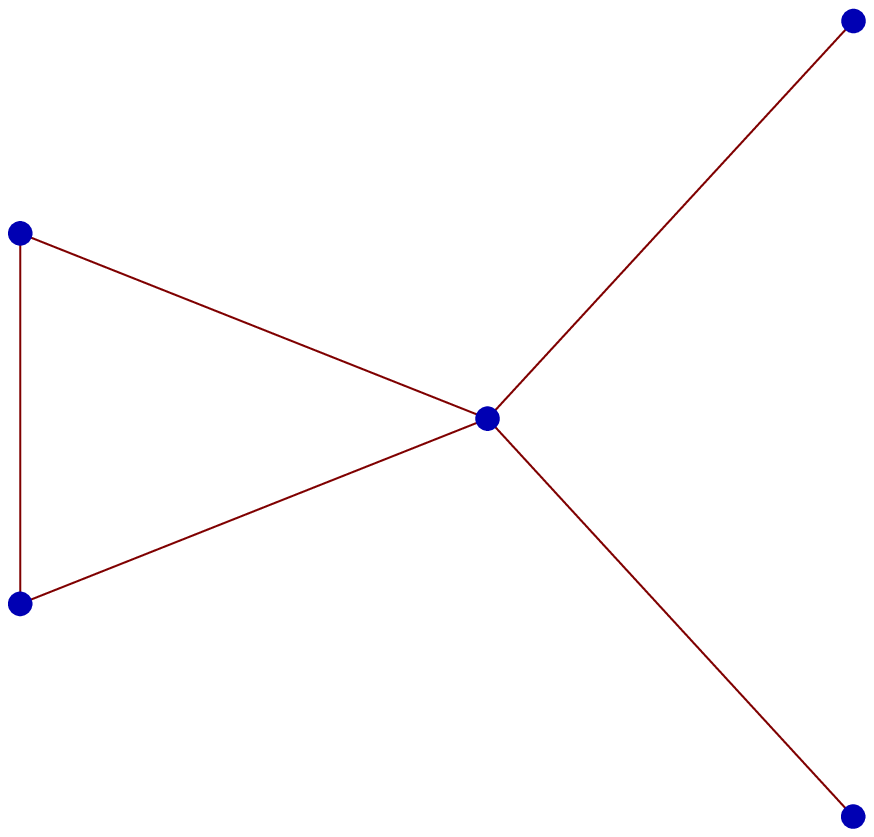} &$58$  &\includegraphics[width=2cm]{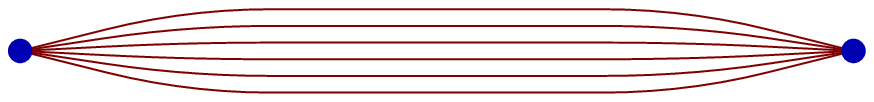} \\
\hline\\
$35$  &\includegraphics[width=2cm]{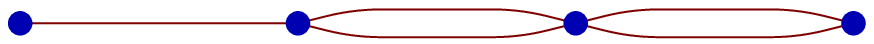} &$43$
&\includegraphics[width=2cm]{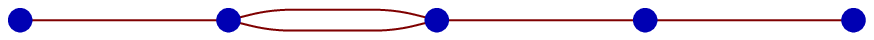}
 &$51$  &\includegraphics[width=2cm]{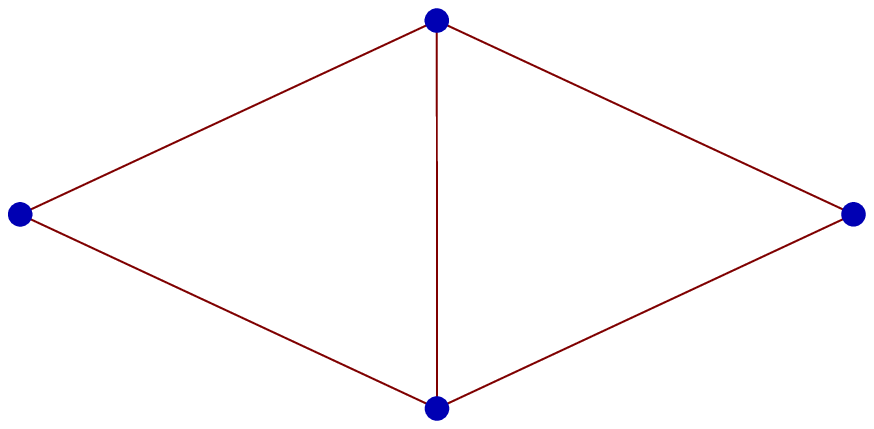} &$59$  &\includegraphics[width=2cm]{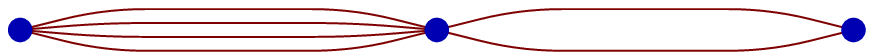}\\
\hline\\
$36$  &\includegraphics[width=2cm]{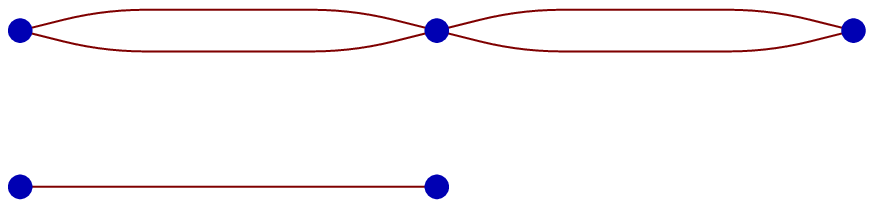}&$44$
&\includegraphics[width=2cm]{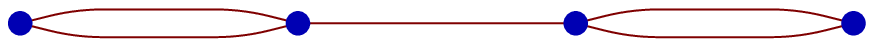}
 &$52$  &\includegraphics[width=2cm]{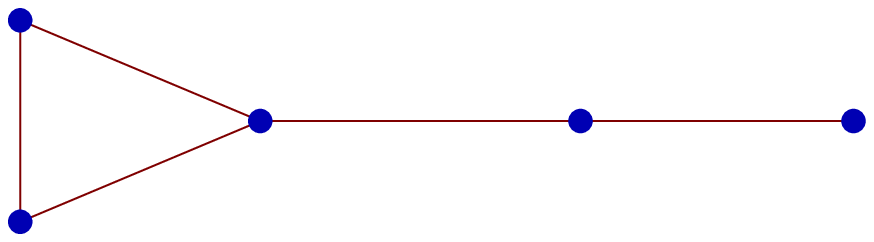} &$60$  &\includegraphics[width=2cm]{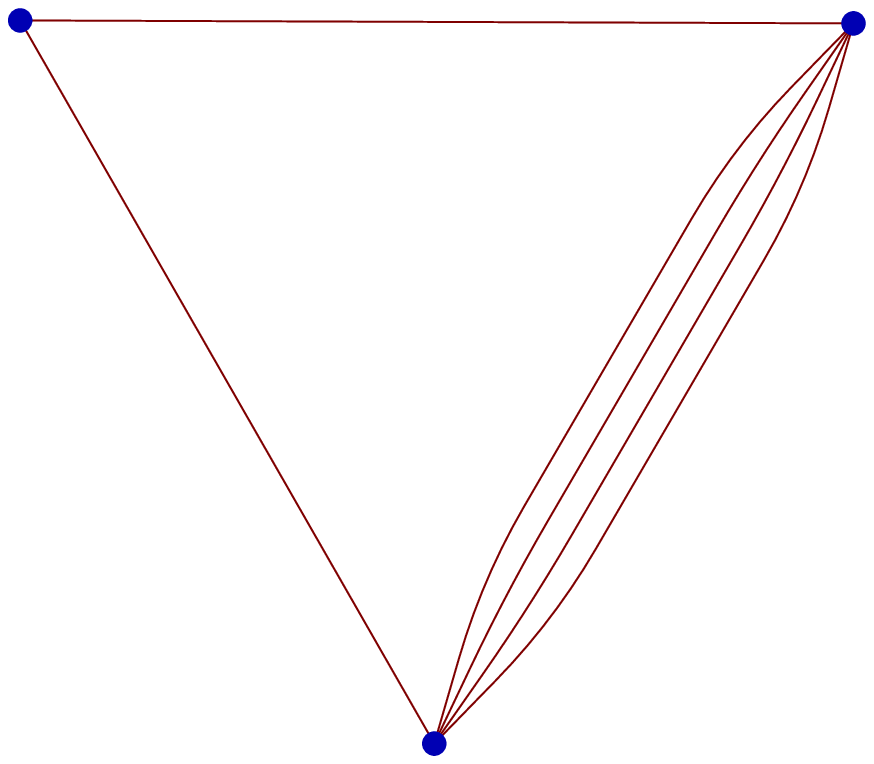}  \\
\hline\\
$37$  &\includegraphics[width=2cm]{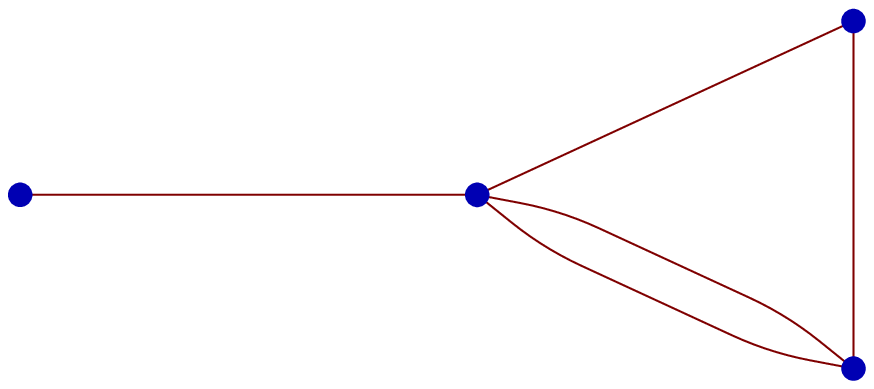}&$45$
&\includegraphics[width=2cm]{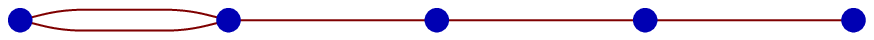}
 &$53$  &\includegraphics[width=2cm]{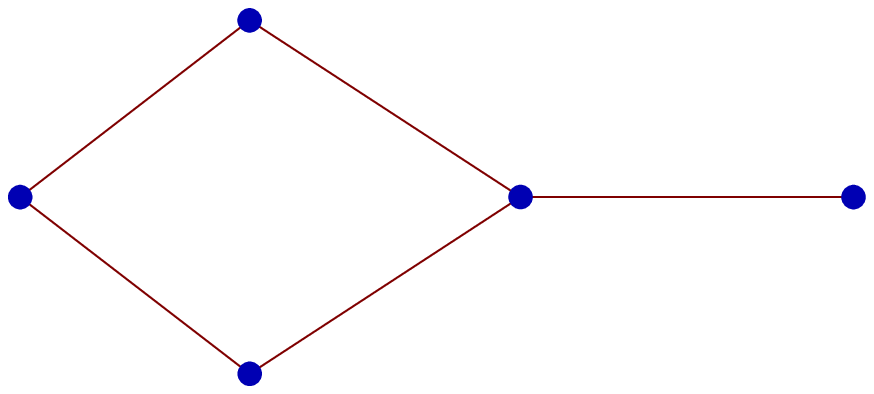} &$61$  &\includegraphics[width=2cm]{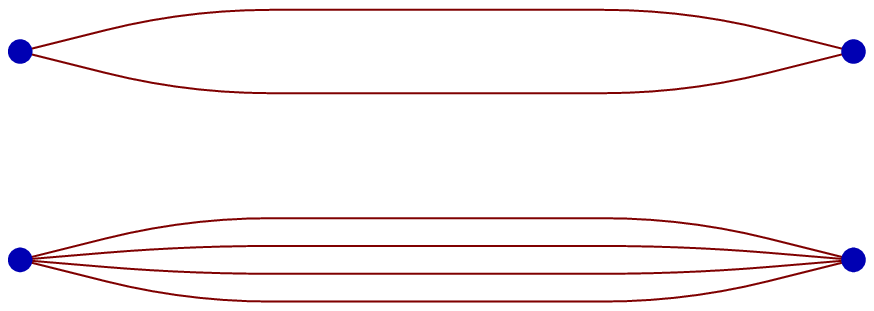} \\
\hline\\
$38$  &\includegraphics[width=2cm]{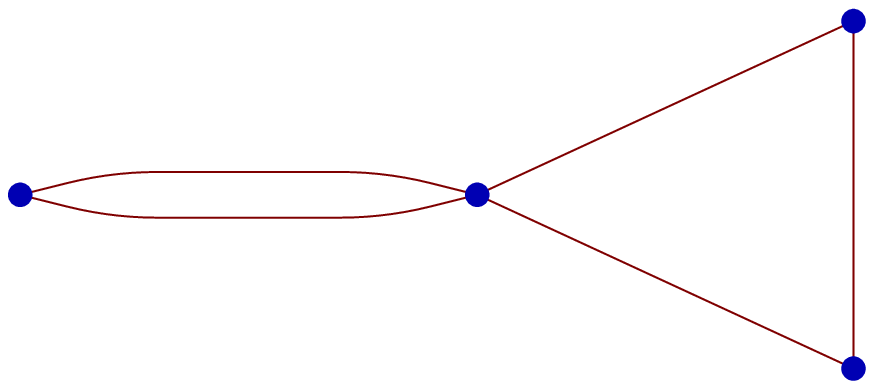}&$46$
&\includegraphics[width=2cm]{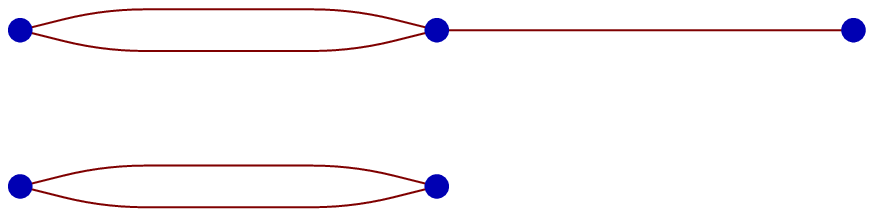}
 &$54$  &\includegraphics[width=2cm]{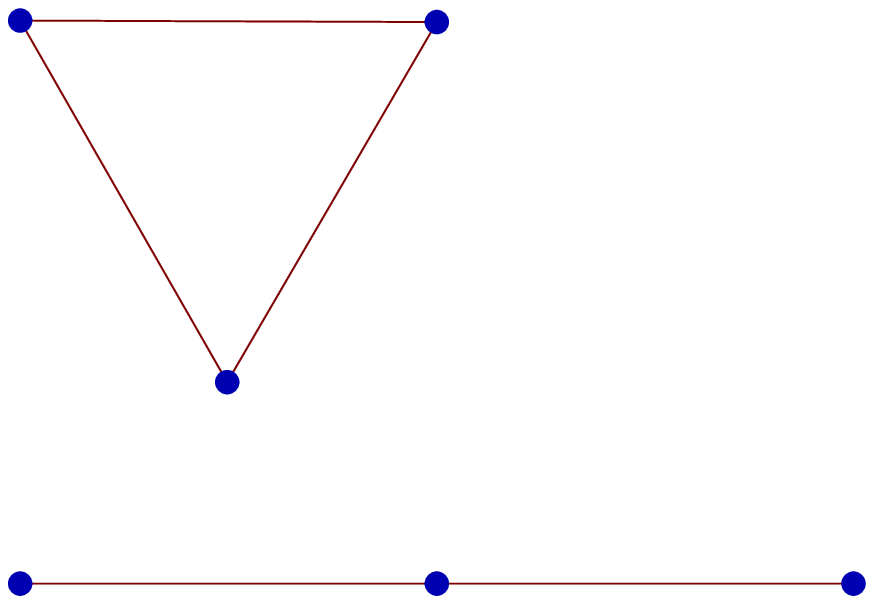} &$62$  &\includegraphics[width=2cm]{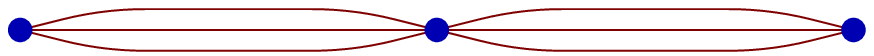}\\
\hline\\
$39$  &\includegraphics[width=2cm]{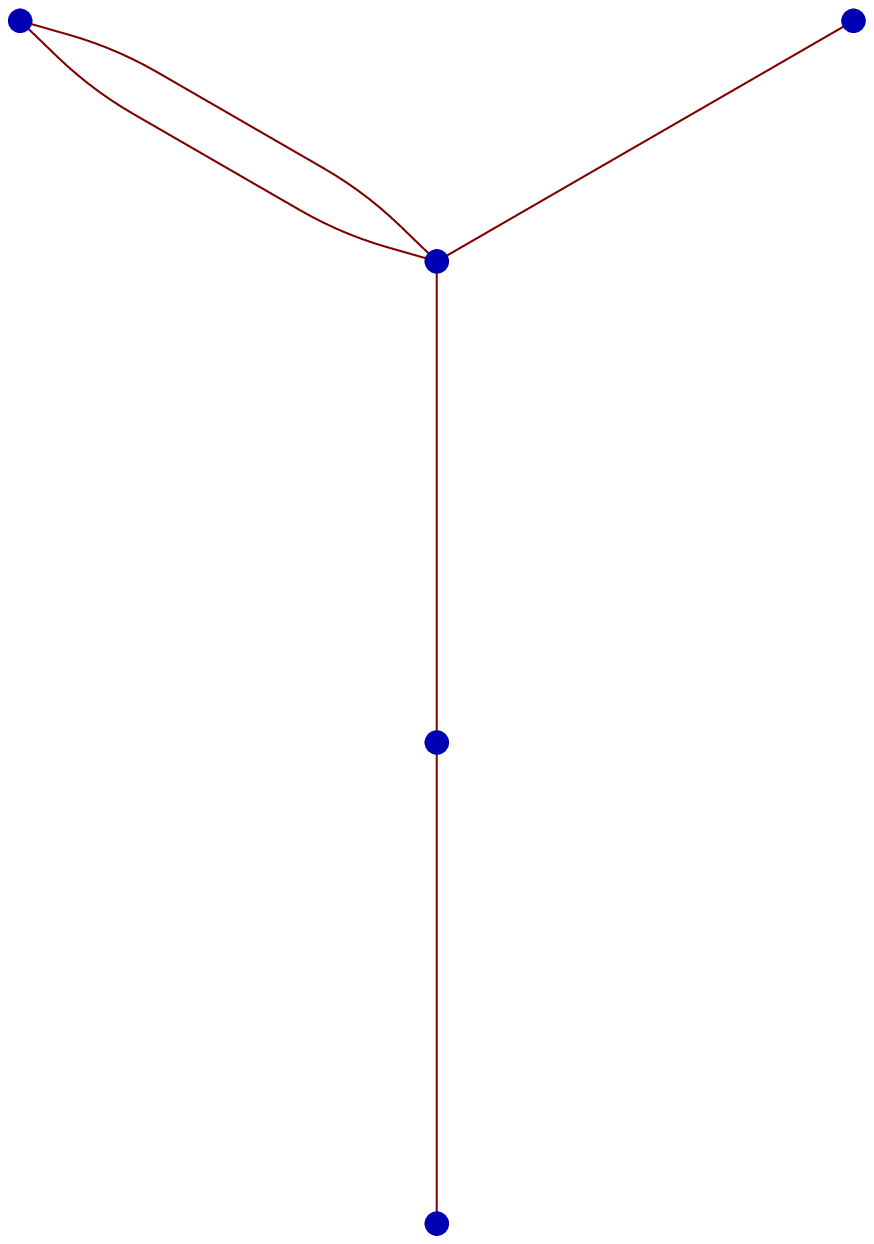}&$47$
&\includegraphics[width=2cm]{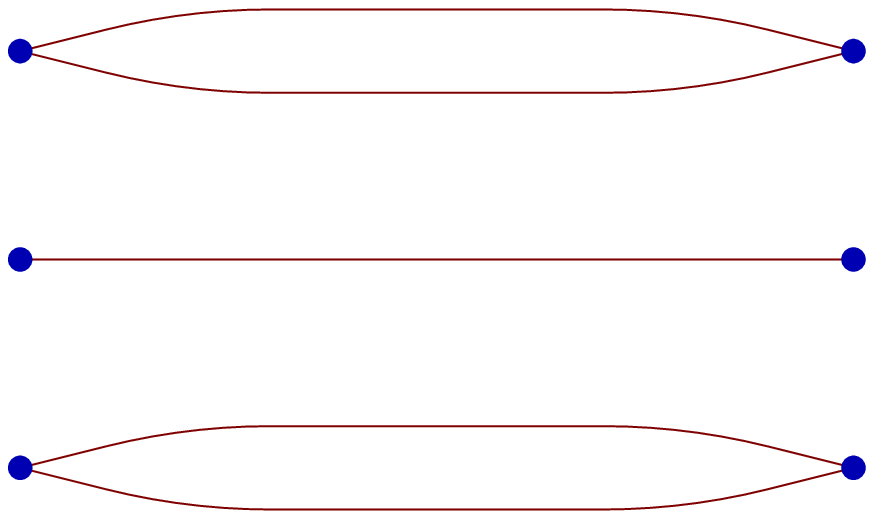}
 &$55$  &\includegraphics[width=2cm]{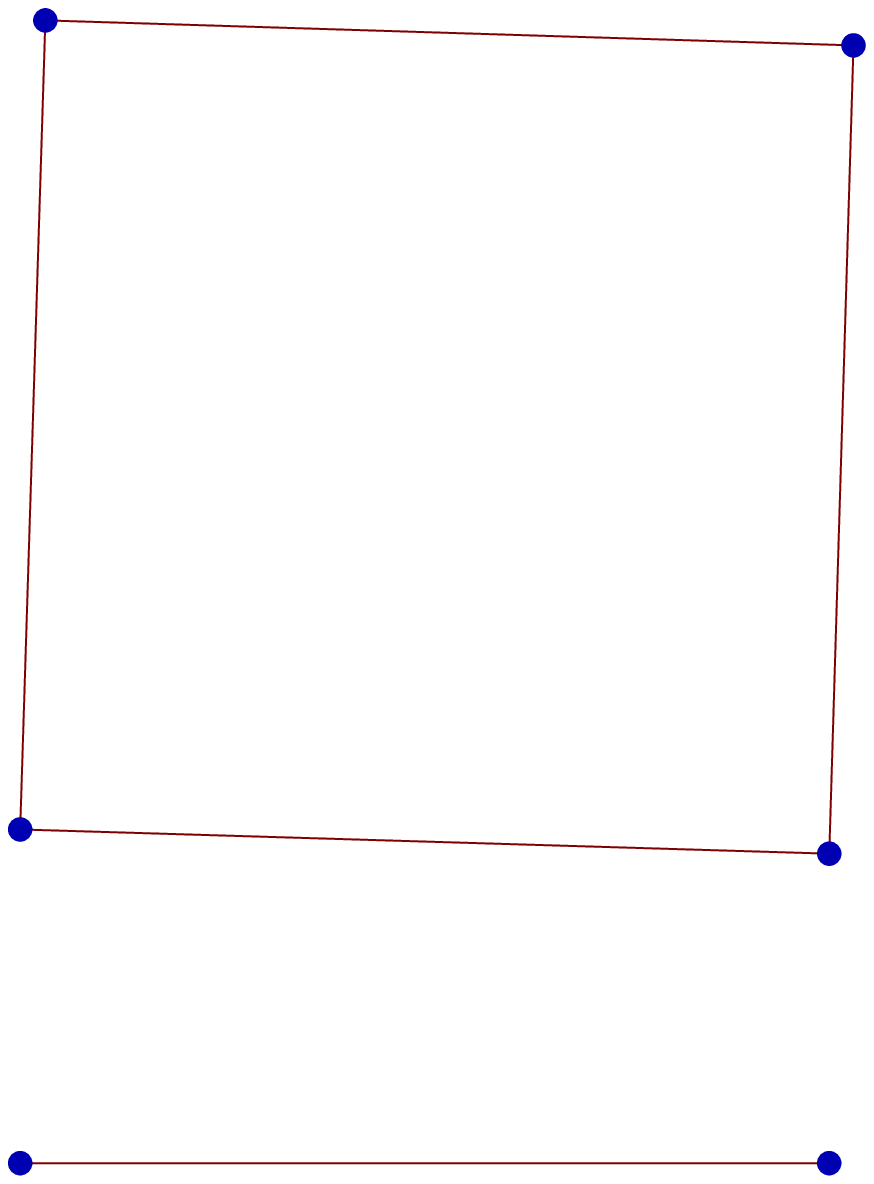} &$63$  &\includegraphics[width=2cm]{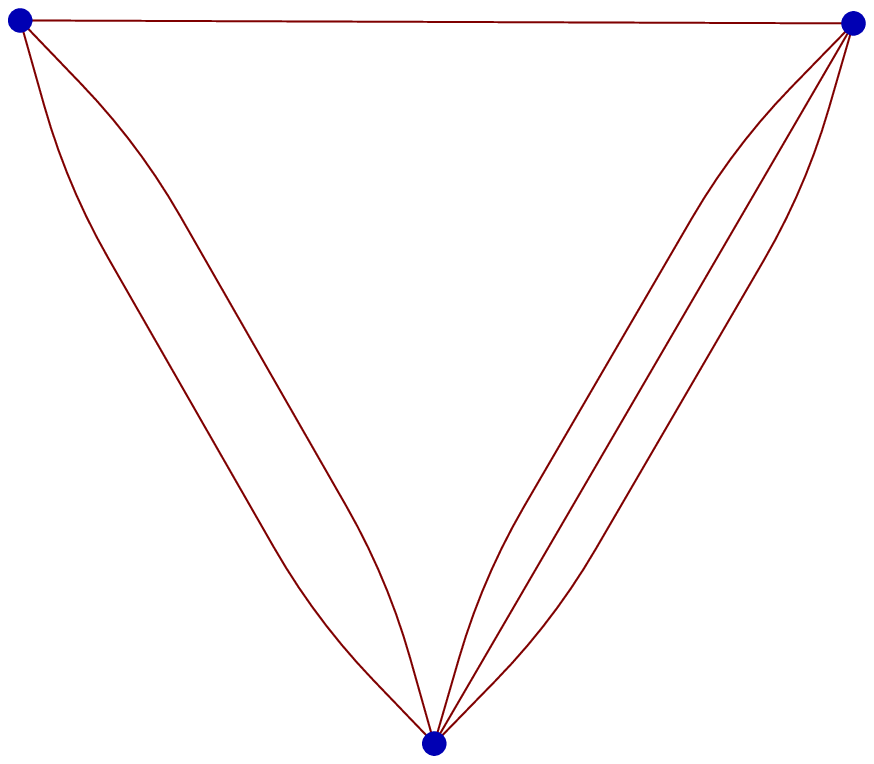}  \\
\hline\\
$40$  &\includegraphics[width=2cm]{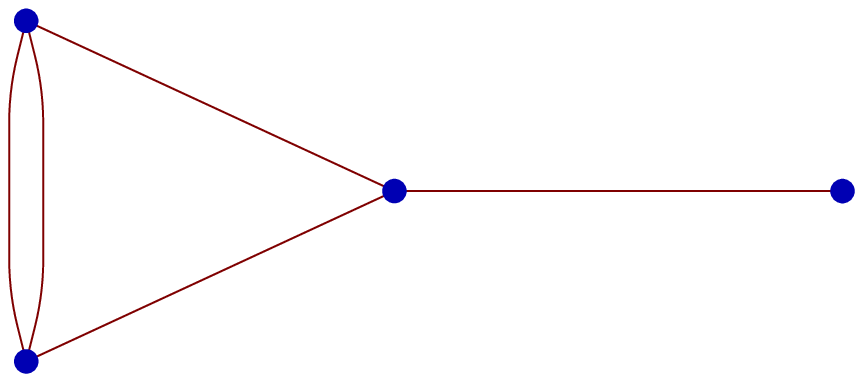} &$48$
&\includegraphics[width=2cm]{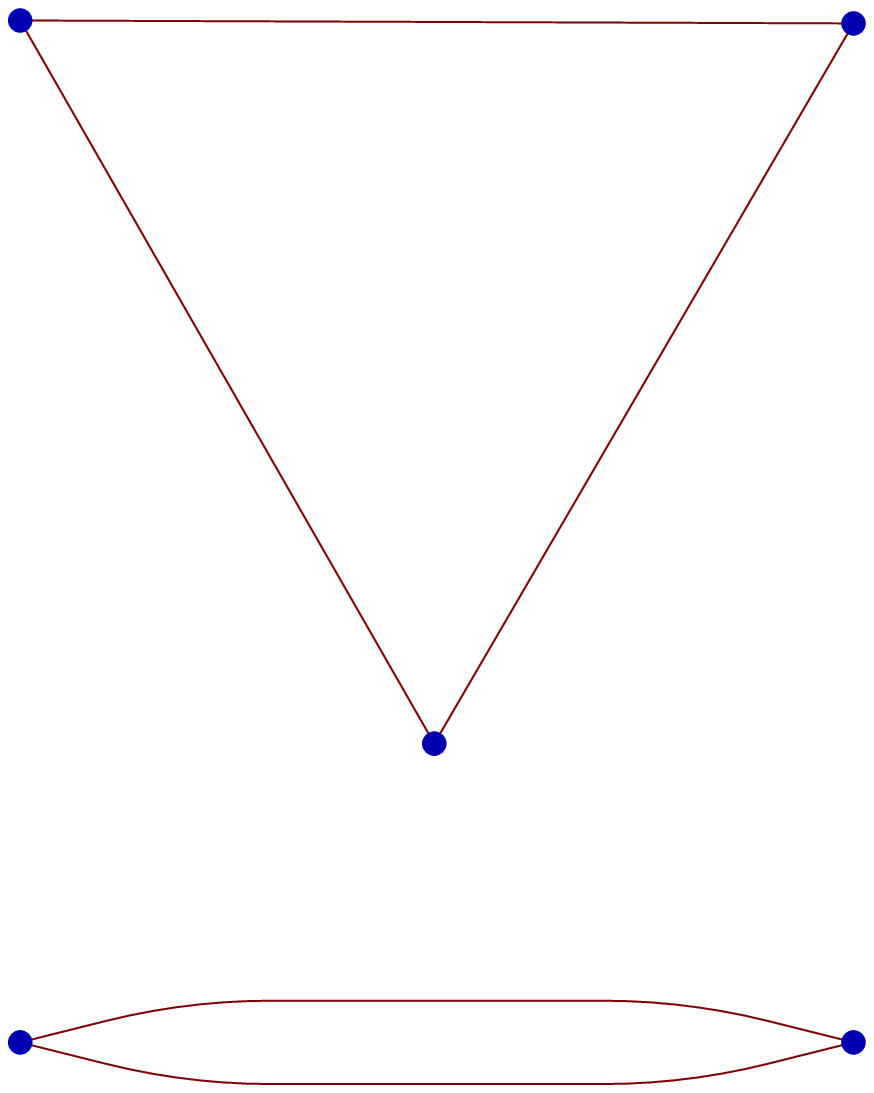}
 &$56$  &\includegraphics[width=2cm]{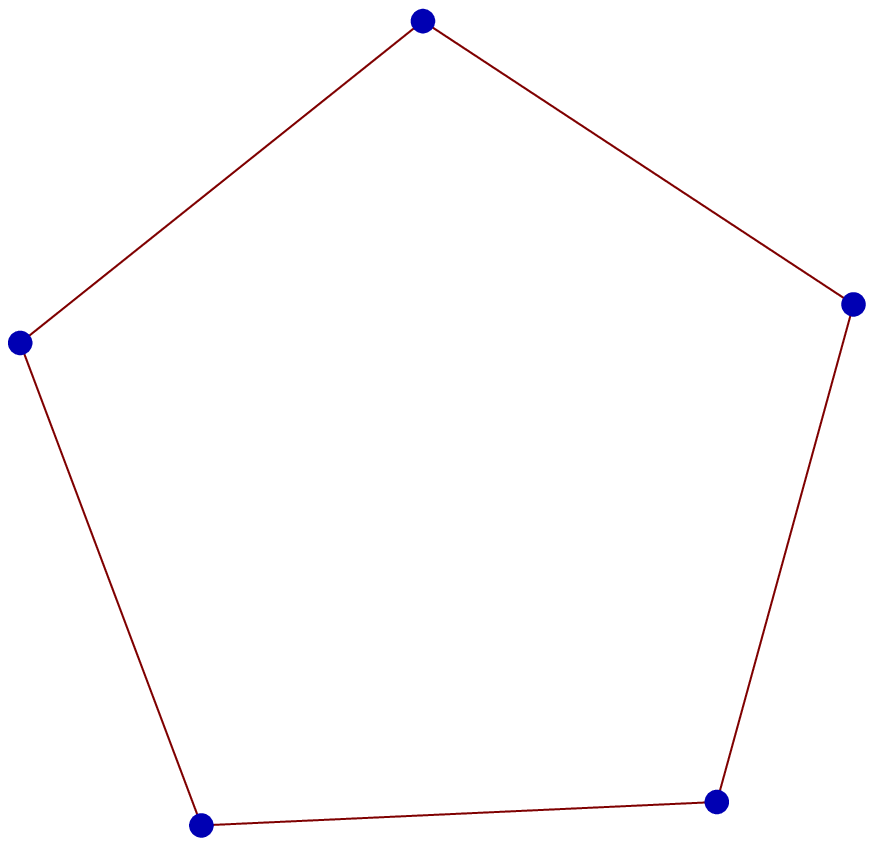} &$64$  &\includegraphics[width=2cm]{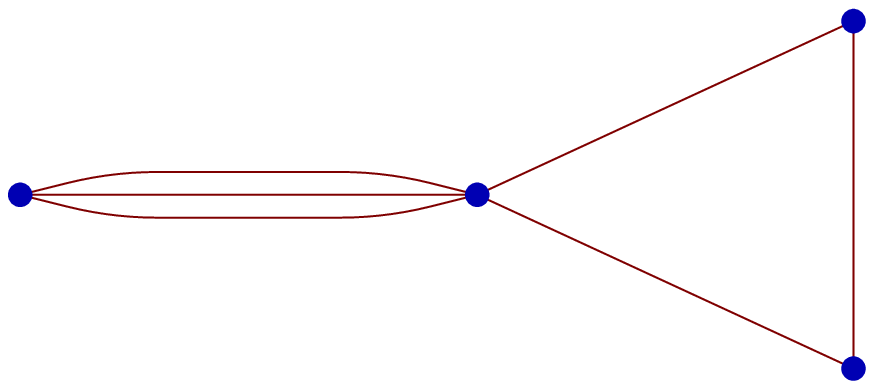} \\
\hline
\end{tabular}
\end{center}
\end{table}

\begin{table}
\begin{center}
\begin{tabular}{||l|l||l|l||l|l||l|l||}
\hline \\
$\nu$ & ${\mathcal G}_\nu$ &$\ldots$ & $\ldots$ &$\ldots$ & $\ldots$ &$\ldots$ & $\ldots$\\
\hline
\hline\\
 $65$  &\includegraphics[width=2cm]{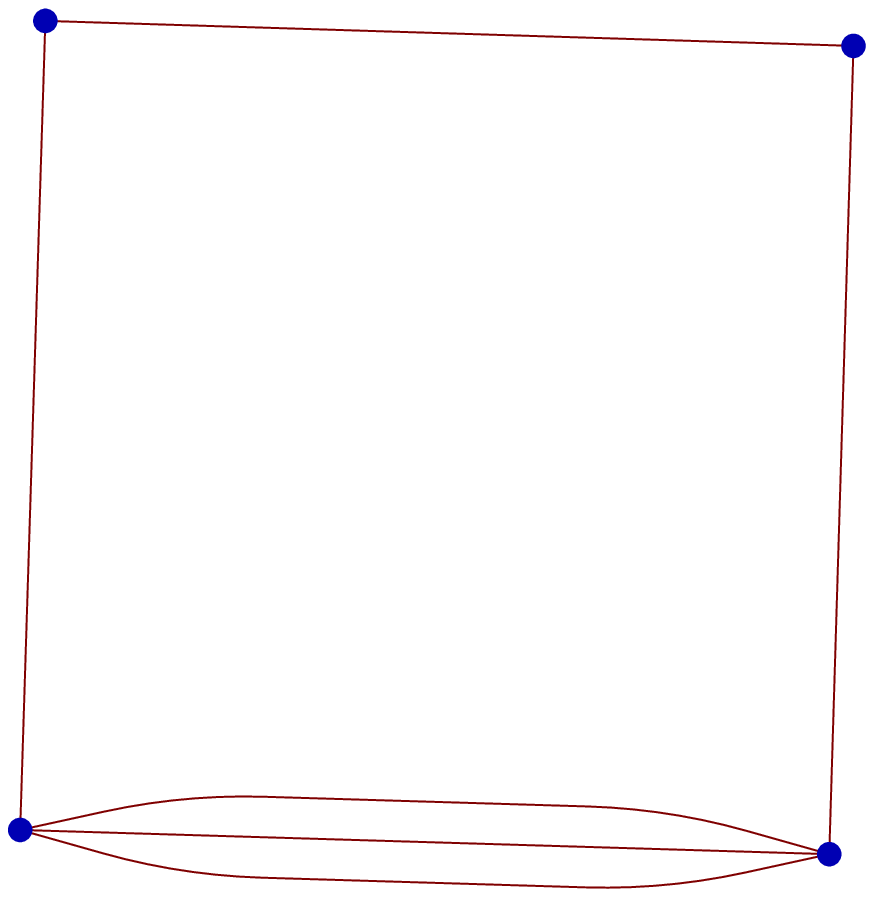} &$71$  &\includegraphics[width=2cm]{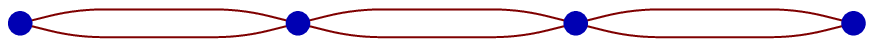}
 &$77$  &\includegraphics[width=2cm]{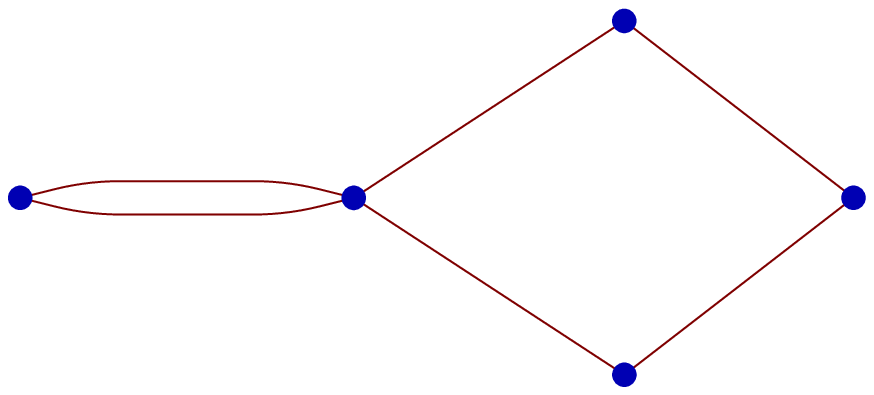} &$83$  &\includegraphics[width=2cm]{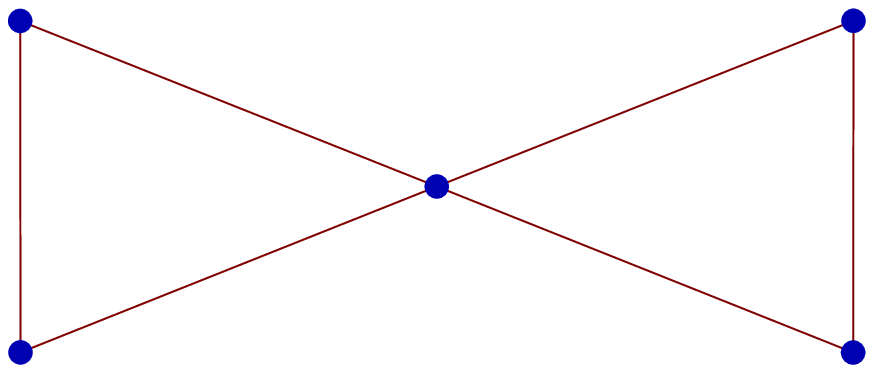}\\
\hline\\
$66$  &\includegraphics[width=2cm]{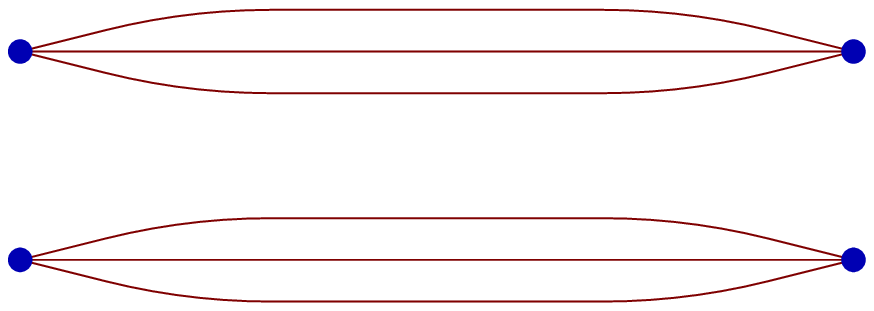} &$72$
&\includegraphics[width=2cm]{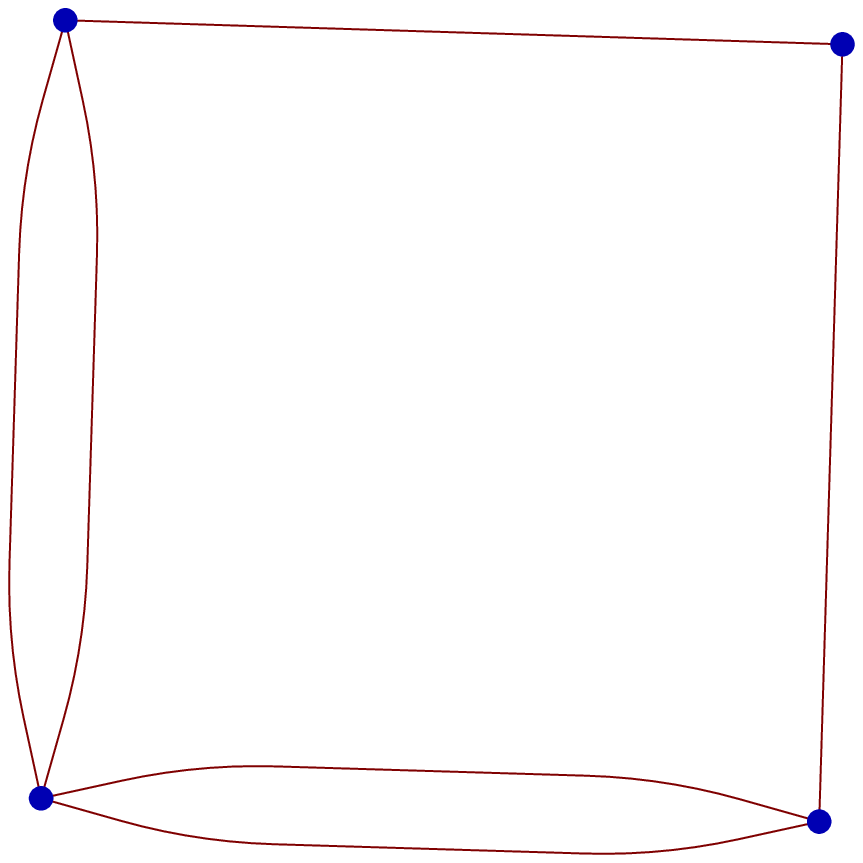}
 &$78$  &\includegraphics[width=2cm]{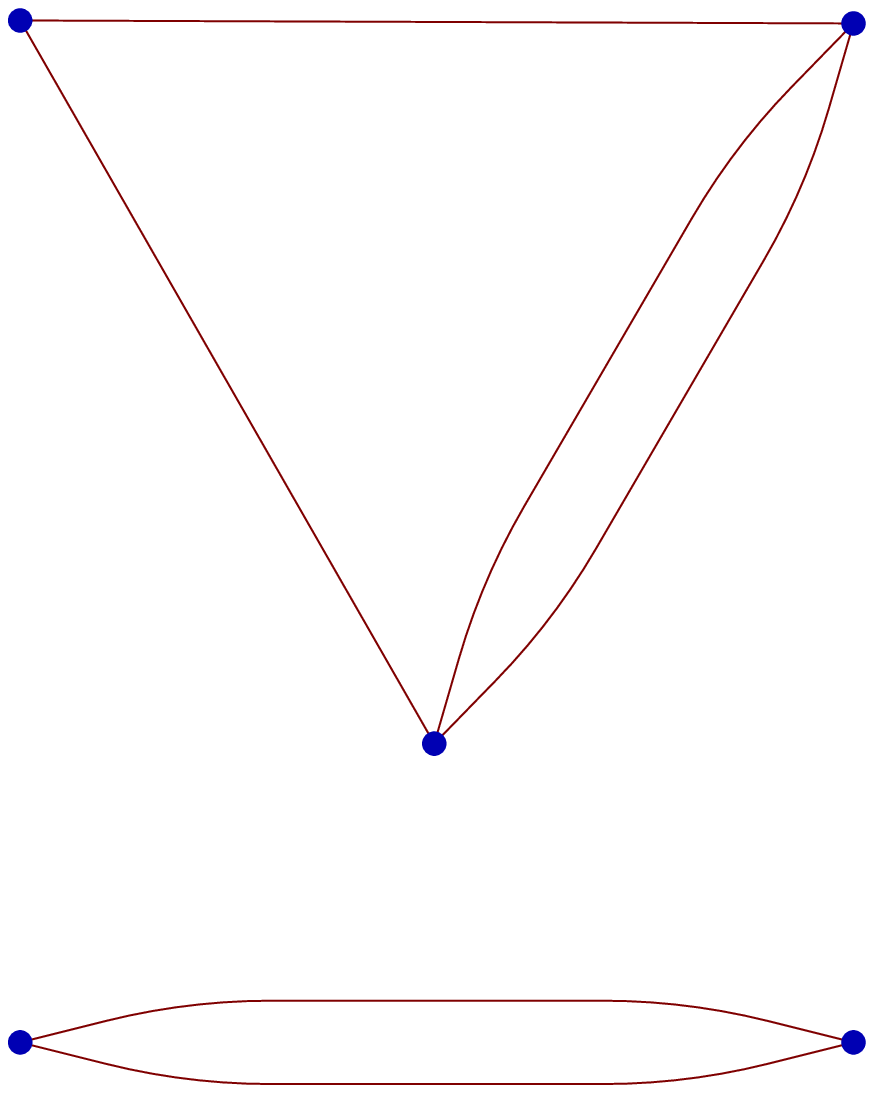} &$84$  &\includegraphics[width=2cm]{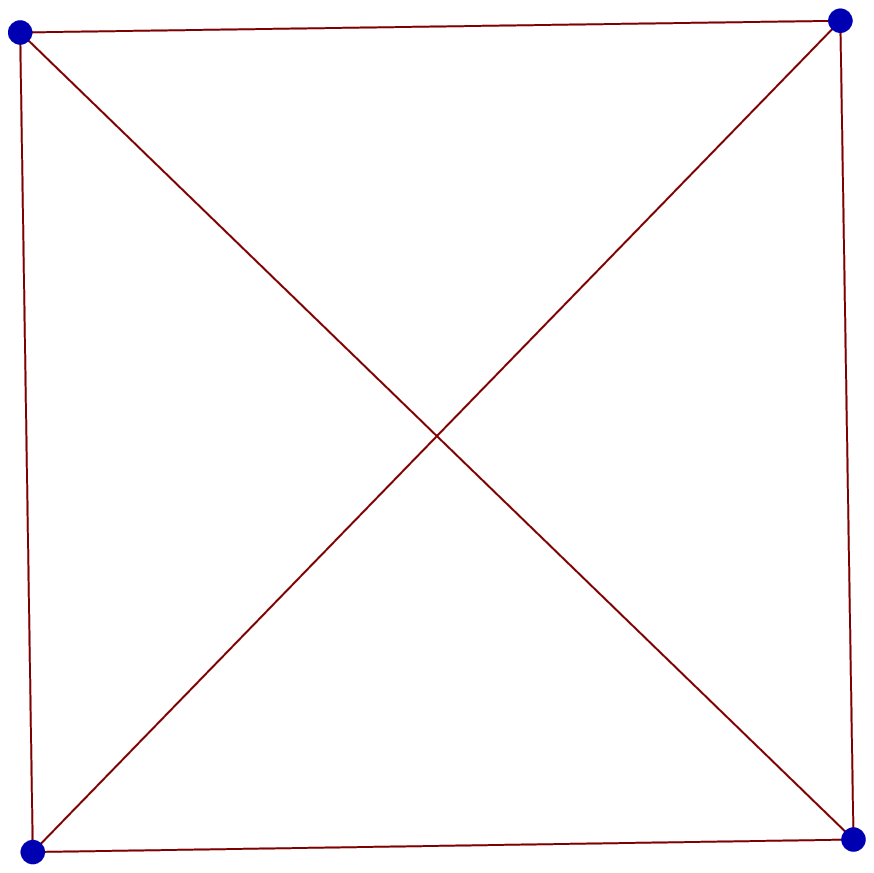} \\
\hline\\
$67$  &\includegraphics[width=2cm]{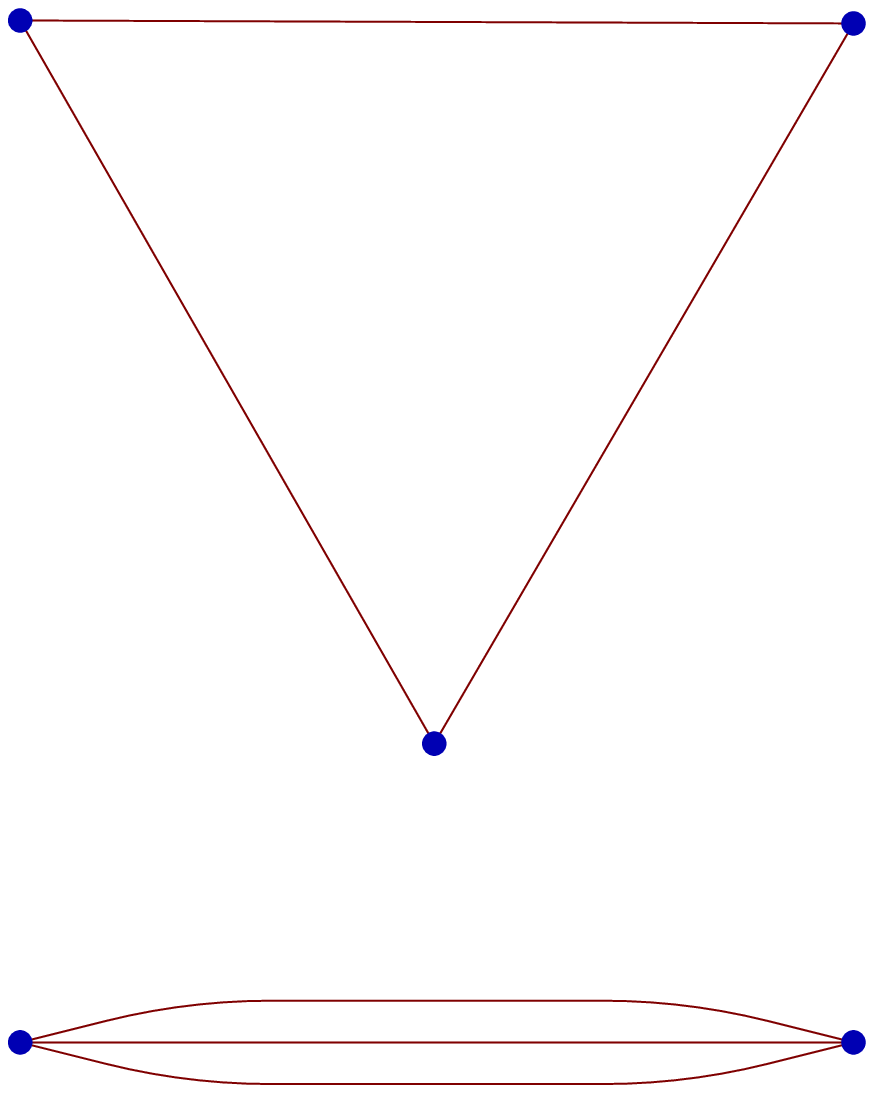} &$73$
&\includegraphics[width=2cm]{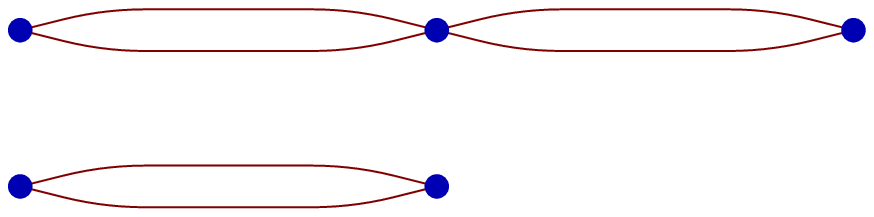}
 &$79$  &\includegraphics[width=2cm]{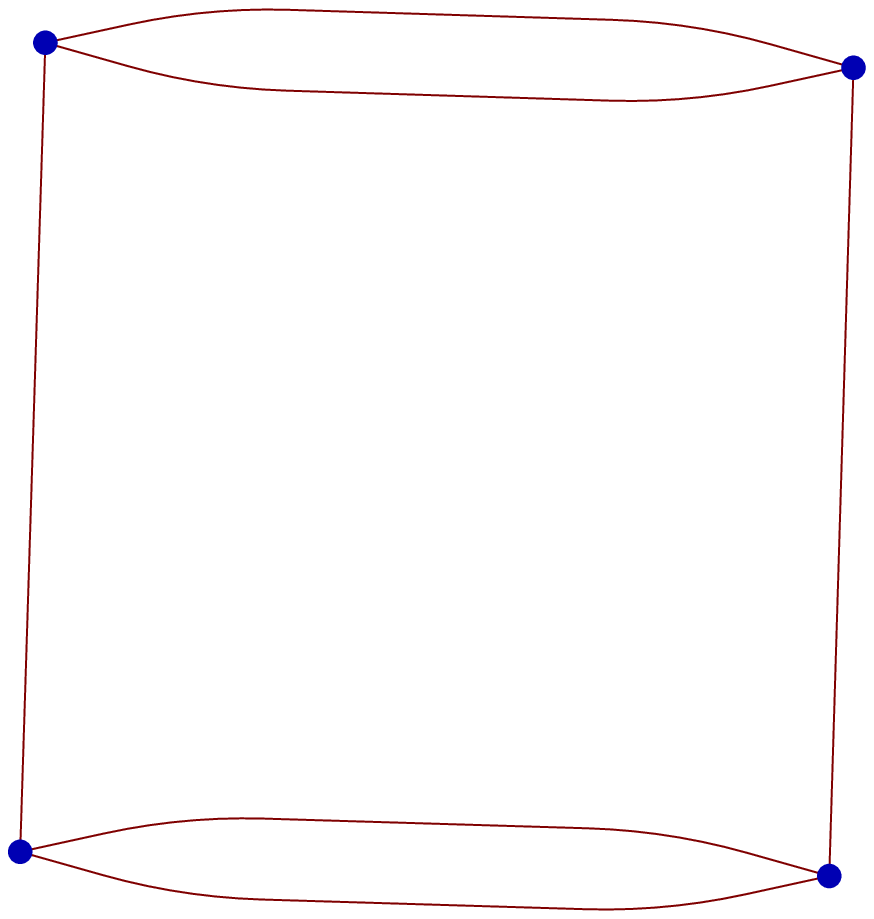} &$85$  &\includegraphics[width=2cm]{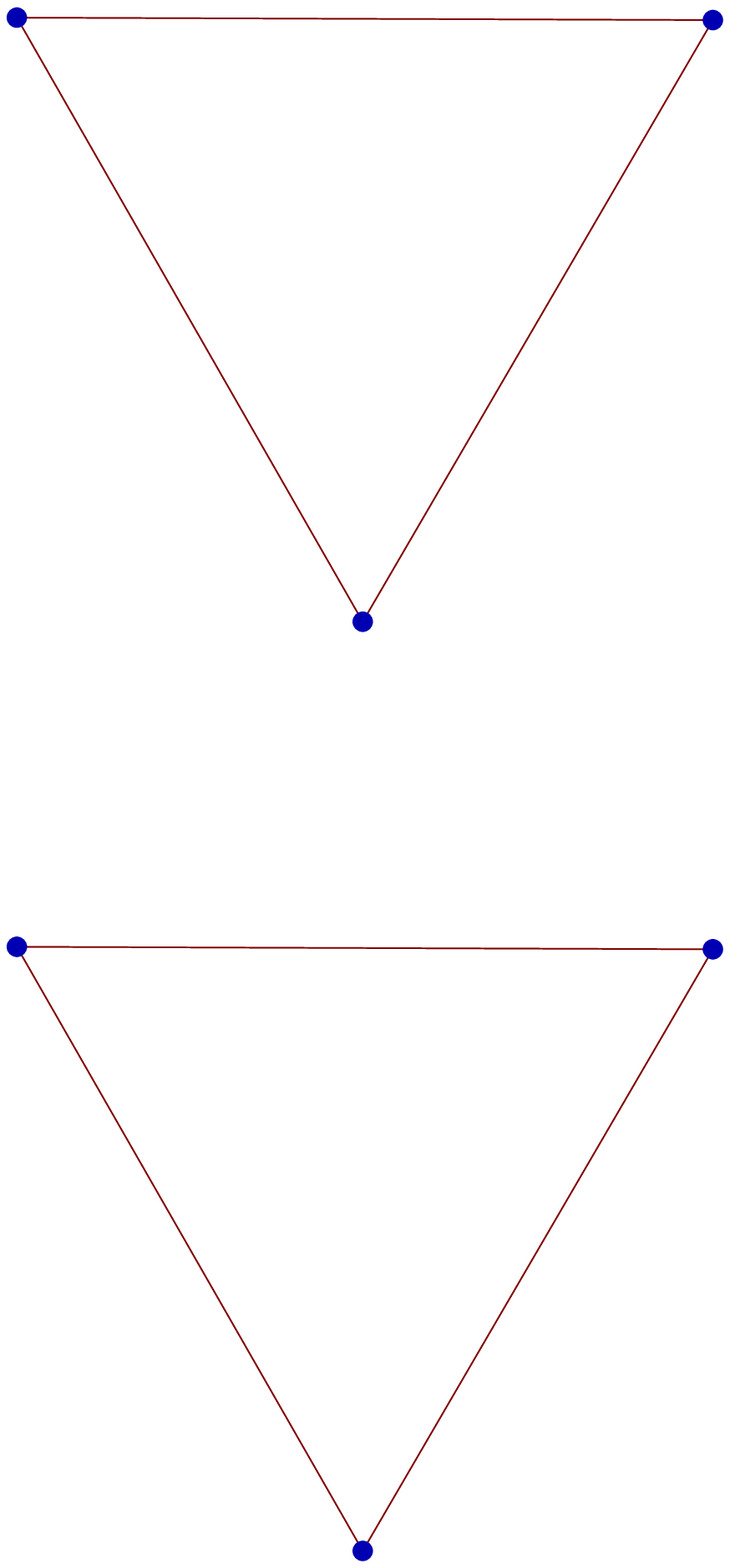}\\
\hline\\
$68$  &\includegraphics[width=2cm]{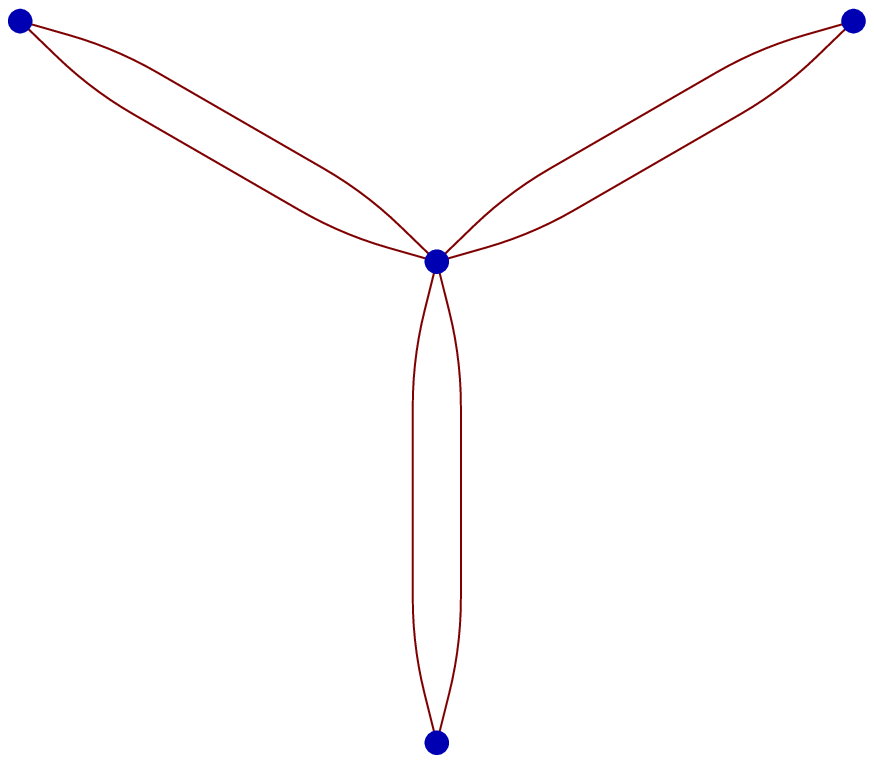}&$74$
&\includegraphics[width=2cm]{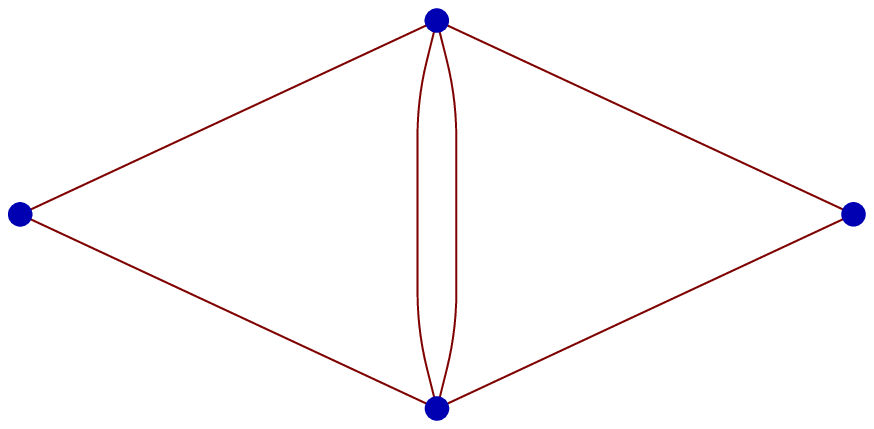}
 &$80$  &\includegraphics[width=2cm]{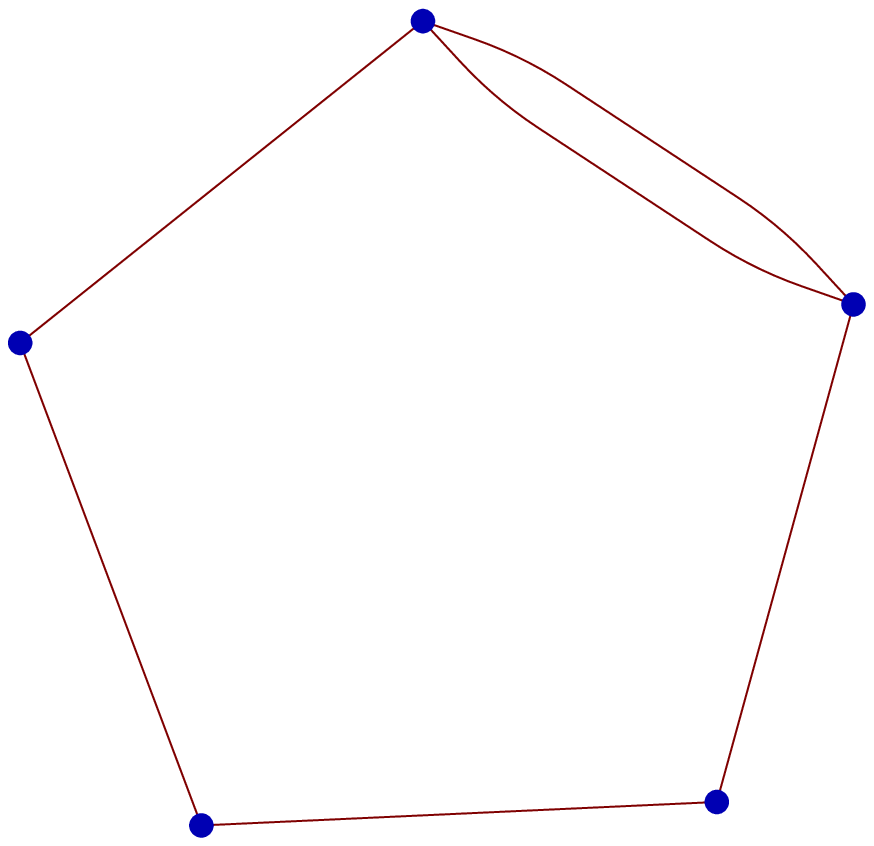} &$86$  &\includegraphics[width=2cm]{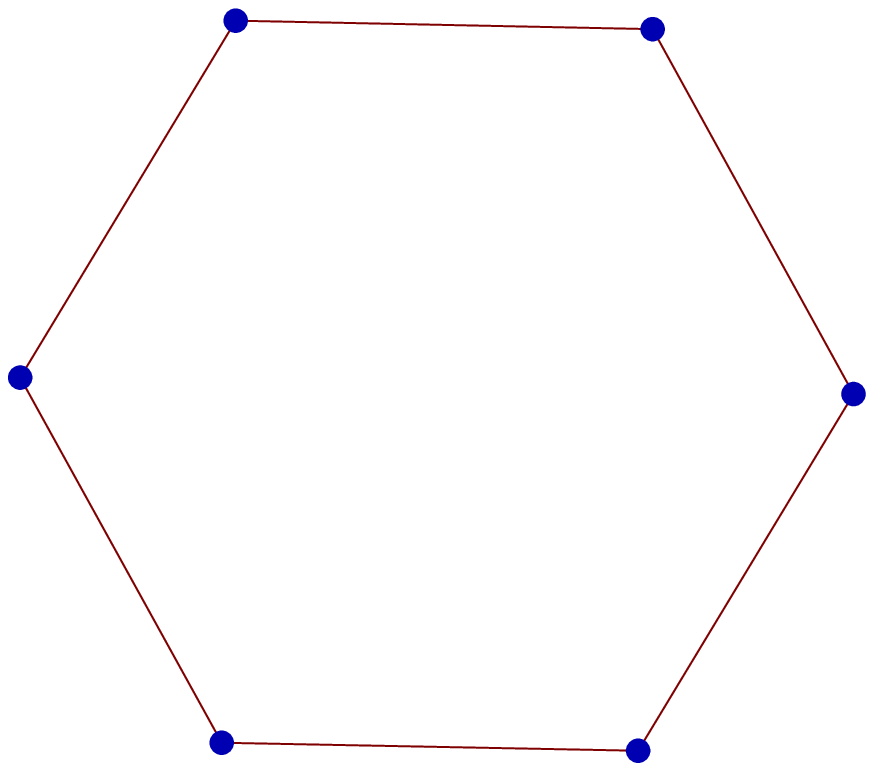}  \\
\hline\\
$69$  &\includegraphics[width=2cm]{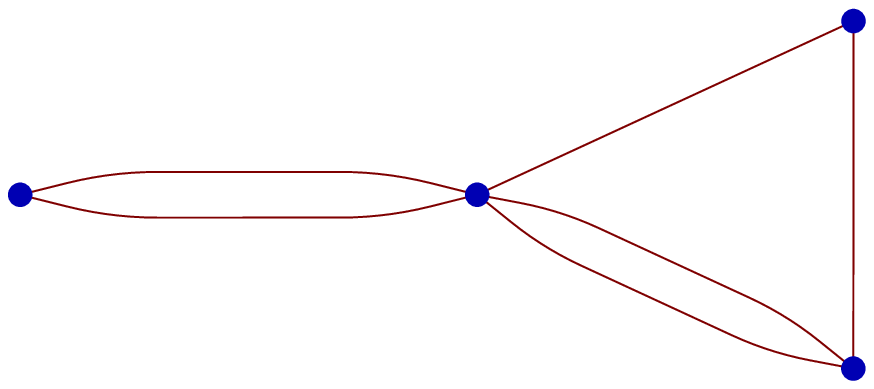}&$75$
&\includegraphics[width=2cm]{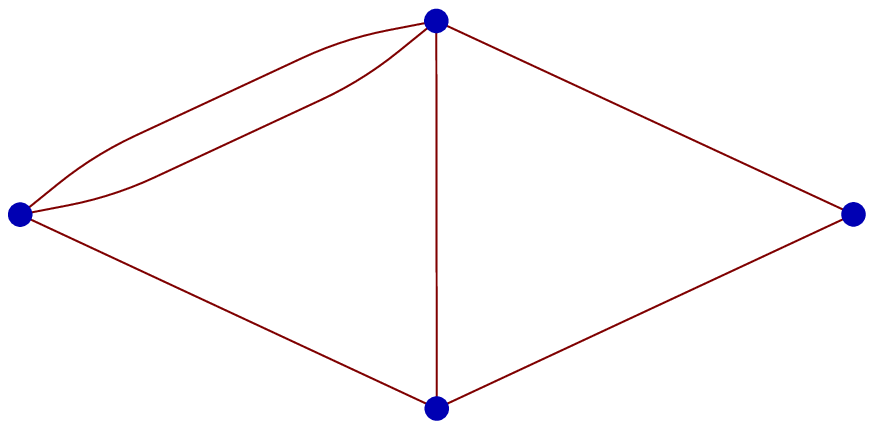}
 &$81$  &\includegraphics[width=2cm]{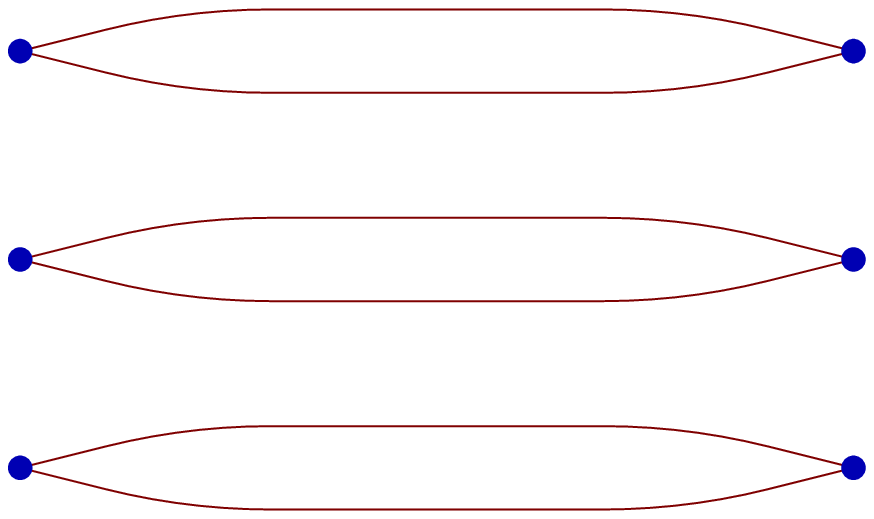} && \\
\hline\\
$70$  &\includegraphics[width=2cm]{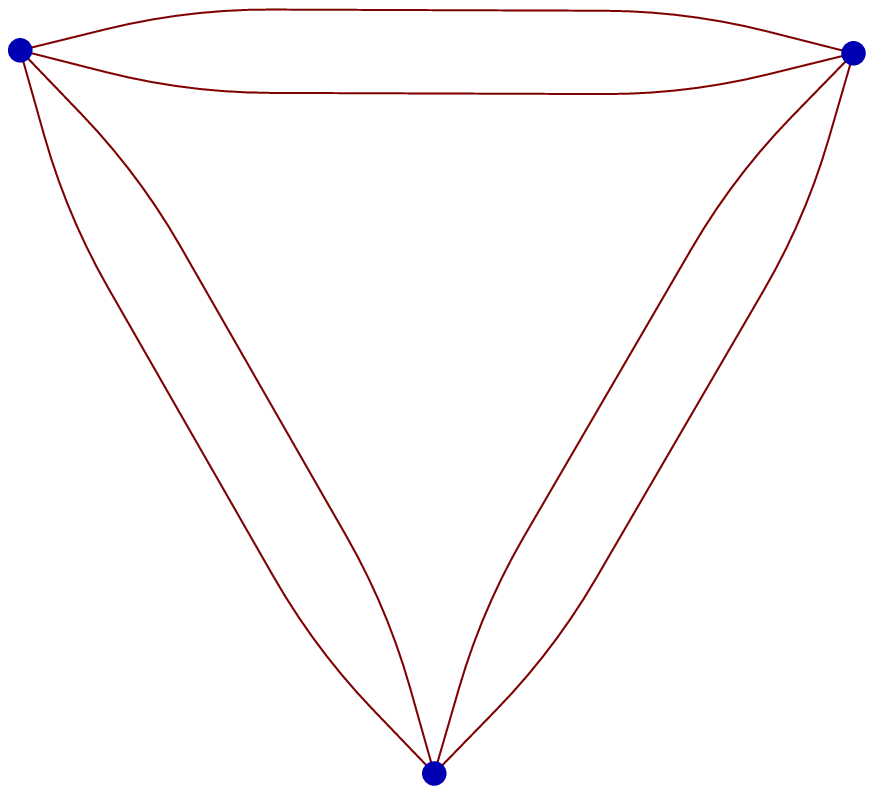}&$76$
&\includegraphics[width=2cm]{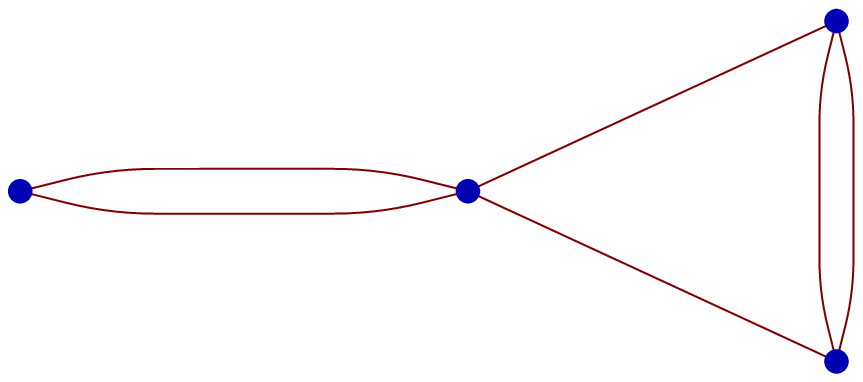}
 &$82$  &\includegraphics[width=2cm]{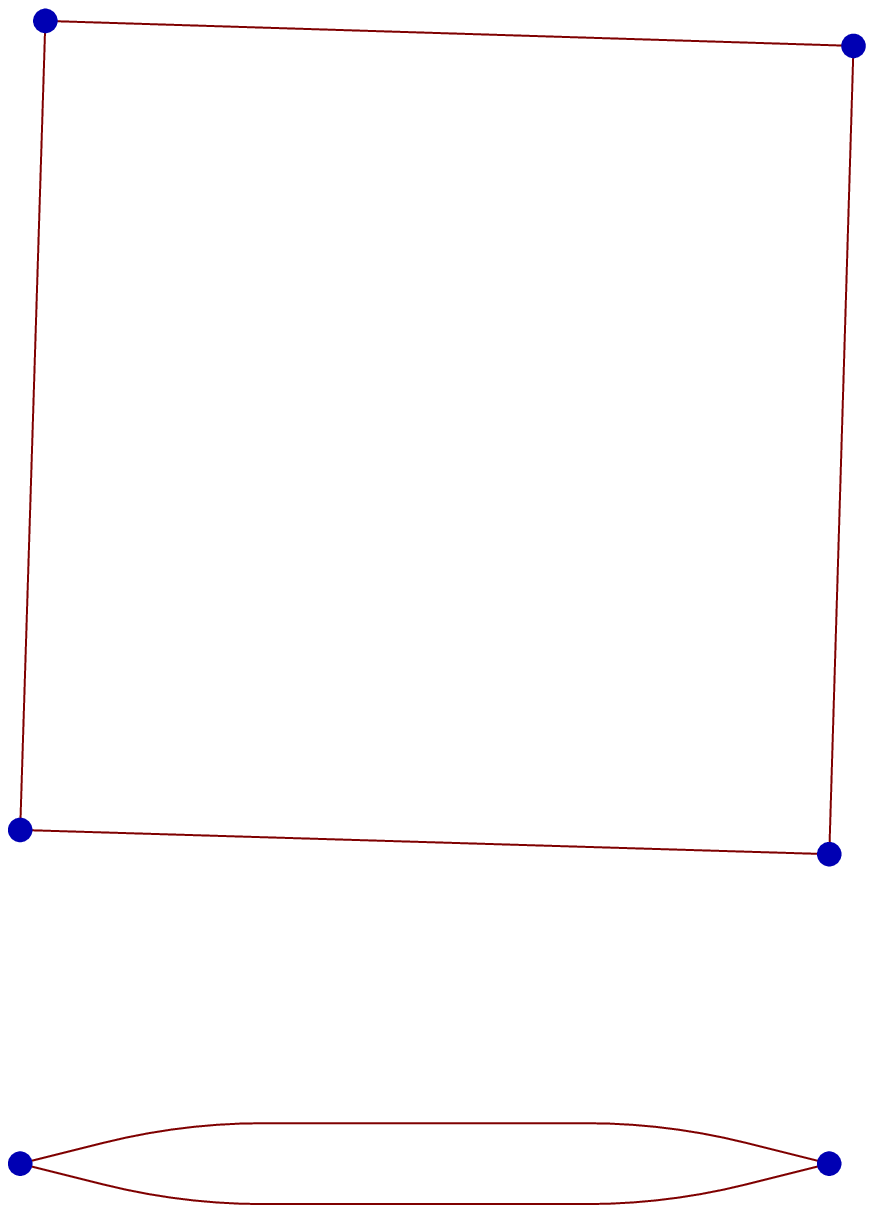} &&\\
\hline
\end{tabular}
\end{center}
\end{table}

In order to write our results in a compact way we will utilize some
graph theoretic notations. This is a common practice when dealing with series expansions,
see e.~g.~\cite{OHZ06}.
 Let ${\mathcal G}$ be a
\underline{multigraph} consisting of $g$ nodes (vertices) and
${\mathcal N}(i,j)={\mathcal N}(j,i)$ bonds (edges) between the
$i$th and the $j$th node. We do not consider ``loops",
i.~e.~${\mathcal N}(i,i)=0$ for all $i=1,\ldots,g$. The total number
of all bonds, $\gamma({\mathcal G})=\sum_{i<j}{\mathcal N}(i,j)$
will be called the \underline{size} of ${\mathcal G}$. ${\mathcal
G}$ is not necessarily connected, see the examples of multigraphs
below. We will identify the set of $g$ nodes with $\{1,2,\ldots,g\}$
and the set of $N$ spins with $\{1,2,\ldots,N\}$.
All multigraphs ${\mathcal G}_\nu,\;\nu=1,\ldots,86$ needed in this paper are represented in table \ref{table1}.\\

From the expansion
\begin{equation}\label{dexp}
\mbox{Tr }H^n=\sum_{\mu_1<\nu_1,\ldots,\mu_n<\nu_n}\;\prod_{i}J_{\mu_i \nu_i}\;\mbox{ Tr }
\left(\prod_{i}\bf{s}_{\mu_i}\cdot\bf{s}_{\nu_i}\right)
\end{equation}
it is clear that the expressions for the moments $t_n$ involve various products of coupling constants $J_{\mu\nu}$.
The structure of these products can be represented by the multigraphs ${\mathcal G}$ defined above, such that the factors
$J_{\mu\nu}^\ell$ correspond to the bonds of ${\mathcal G}$ with multiplicity $\ell$.
The sum of different products in (\ref{dexp}) of the same structure
will be obtained by an \underline{evaluation} of ${\mathcal G}$, denoted by $\overline{\mathcal G}$, for the spin system
under consideration. $\overline{\mathcal G}$ denotes a real number which depends on the coupling constants and only implicitly on
the number $N$ of spins.
This number will be defined according to the following statements.
\begin{enumerate}
\item If $g>N$ we set $\overline{\mathcal G}=0$.\\
\item If $g\le N$ we consider the \underline{symmetry group} $G$ of ${\mathcal G}$, defined by
\begin{eqnarray} \nonumber
G&=&\{\pi\in{\mathcal S}_N | \pi \mbox{ maps } \{1,2,\ldots,g\} \mbox{ onto } \{1,2,\ldots,g\} \mbox{ and }\\
\label{d1}
&&{\mathcal N}(i,j)={\mathcal N}(\pi(i),\pi(j)) \mbox{ for all } 1\le i,j \le g
\}
\end{eqnarray}
Further we consider the quotient set ${\mathcal S}_N / G$ consisting of left cosets
$\pi\,G,\;\pi\in{\mathcal S}_N $ and choose a representative $\pi_\ell$ from each coset,
$\pi_\ell\in \pi\,G, \quad\ell=1,\ldots,L\equiv \left| {\mathcal S}_N / G \right| = \frac{N!}{|G|}$.\\
\item Then we define
\begin{equation}\label{d2}
\overline{\mathcal G}\equiv \sum_{\ell=1}^L\, \prod_{1\le i<j\le
g} \, \left( J_{\pi_\ell(i),\pi_\ell(j)} \right)^{{\mathcal
N}(i,j)} \;.
\end{equation}
\end{enumerate}
Obviously, the definition of $\overline{\mathcal G}$ does not depend on the choice of representatives $\pi_\ell$
since the product $\prod_{1\le i<j\le g} \,\left(J_{\pi_\ell(i),\pi_\ell(j)}\right)^{{\mathcal N}(i,j)}$ is invariant under
permutations from the symmetry group $\pi\in G$.\\

In order to illustrate this definition we consider an example of $N=4$ spins
and ${\mathcal G}=\ti$, hence $g=3<4=N$. The symmetry group $G$ consists of all permutations
of $\{1,2,3,4\}$ leaving $4$ fixed, hence $|G|=3!=6$. Thus we have $L=\frac{4!}{3!}=4$ left cosets
from which we choose the representatives
$\pi_1=(1)(2)(3)(4)=\mbox{id},\;\pi_2=(1234),\;\pi_3=(13)(24),\;\pi_4=(1432)$.
Hence $\overline{\mathcal G}=J_{12}J_{23}J_{13}+J_{23}J_{34}J_{23}+J_{34}J_{14}J_{13}+J_{14}J_{12}J_{24}$.\\

The coefficients $c_n$ of the susceptibility's HTE (and similarly the $a_n$ of the free energy HTE)
will contain products of evaluations
$\overline{{\mathcal G}_\nu}\;\overline{{\mathcal G}_\mu}$. These expressions can be simplified using
rules which transform such products into linear combinations of other evaluations. To give an example, we
consider
$\overline{{\mathcal G}_1}\;\overline{{\mathcal G}_2}=
\overline{\;\;\di}\;\overline{\;\;\dii}=\left(\sum_{\mu<\nu}J_{\mu\nu}\right)\,\left(\sum_{\kappa<\lambda}J_{\kappa\lambda}^2\right)$.
It is obvious that this product can be written as a sum over evaluations of the three multigraphs which can be
combined from $\;\di\;$ and $\;\dii\;$, namely  $\;\di\dii\;$, $\;\tii\;$ and $\;\diii$. Similar expressions can be derived for other products
of evaluations where suitable coefficients will compensate the change of symmetry groups. For the purpose of the present paper we
will need the following ``product rules":
\begin{eqnarray}\label{d3a}
\overline{{\mathcal G}_1}\;\overline{{\mathcal G}_2}
&=&
\overline{{\mathcal G}_4}+\overline{{\mathcal G}_5}+\overline{{\mathcal G}_6},\\ \label{d3b}
\overline{{\mathcal G}_2}^2
&=&
\overline{{\mathcal G}_9}+2\overline{{\mathcal G}_{12}}+2\overline{{\mathcal G}_{17}},\\ \label{d3c}
\overline{{\mathcal G}_2}\;\overline{{\mathcal G}_3}
&=&
\overline{{\mathcal G}_{10}}+\overline{{\mathcal G}_{13}}+\overline{{\mathcal G}_{14}}
+\overline{{\mathcal G}_{16}}+\overline{{\mathcal G}_{18}},\\ \label{d3d}
\overline{{\mathcal G}_1}\;\overline{{\mathcal G}_4}
&=&
\overline{{\mathcal G}_9}+\overline{{\mathcal G}_{10}}+\overline{{\mathcal G}_{11}},\\ \label{d3e}
\overline{{\mathcal G}_2}\;\overline{{\mathcal G}_4}
&=&
\overline{{\mathcal G}_{23}}+\overline{{\mathcal G}_{26}}+\overline{{\mathcal G}_{31}},\\ \label{d3f}
\overline{{\mathcal G}_1}\;\overline{{\mathcal G}_7}
&=&
\overline{{\mathcal G}_{14}}+\overline{{\mathcal G}_{19}}+\overline{{\mathcal G}_{20}},\\ \label{d3g}
\overline{{\mathcal G}_2}\;\overline{{\mathcal G}_4}
&=&
\overline{{\mathcal G}_{23}}+\overline{{\mathcal G}_{26}}+\overline{{\mathcal G}_{31}},\\ \label{d3h}
\overline{{\mathcal G}_3}\;\overline{{\mathcal G}_4}
&=&
\overline{{\mathcal G}_{24}}+\overline{{\mathcal G}_{27}}+\overline{{\mathcal G}_{28}}
+\overline{{\mathcal G}_{30}}+\overline{{\mathcal G}_{32}},\\ \label{d3i}
\overline{{\mathcal G}_4}^2
&=&
\overline{{\mathcal G}_{58}}+2\overline{{\mathcal G}_{62}}+2\overline{{\mathcal G}_{66}},\\ \label{d3j}
\overline{{\mathcal G}_2}\;\overline{{\mathcal G}_5}
&=&
\overline{{\mathcal G}_{24}}+\overline{{\mathcal G}_{26}}+2\overline{{\mathcal G}_{33}}
+2\overline{{\mathcal G}_{34}}+\overline{{\mathcal G}_{35}}
+2\overline{{\mathcal G}_{44}}+\overline{{\mathcal G}_{46}},\\ \label{d3k}
\overline{{\mathcal G}_2}\;\overline{{\mathcal G}_6}
&=&
\overline{{\mathcal G}_{25}}+\overline{{\mathcal G}_{31}}+2\overline{{\mathcal G}_{35}}
+2\overline{{\mathcal G}_{36}}+\overline{{\mathcal G}_{46}}
+2\overline{{\mathcal G}_{47}},\\ \label{d3l}
\overline{{\mathcal G}_2}\;\overline{{\mathcal G}_7}
&=&
\overline{{\mathcal G}_{28}}+\overline{{\mathcal G}_{38}}+\overline{{\mathcal G}_{48}},\\ \label{d3m}
\overline{{\mathcal G}_3}\;\overline{{\mathcal G}_7}
&=&
\overline{{\mathcal G}_{34}}+\overline{{\mathcal G}_{37}}+\overline{{\mathcal G}_{50}}
+2\overline{{\mathcal G}_{51}}+\overline{{\mathcal G}_{52}}
+\overline{{\mathcal G}_{54}},\\ \label{d3n}
\overline{{\mathcal G}_2}\;\overline{{\mathcal G}_7}
&=&
\overline{{\mathcal G}_{60}}+\overline{{\mathcal G}_{64}}+2\overline{{\mathcal G}_{67}},\\ \label{d3o}
\overline{{\mathcal G}_7}^2
&=&
\overline{{\mathcal G}_{70}}+2\overline{{\mathcal G}_{74}}++2\overline{{\mathcal G}_{83}}
+2\overline{{\mathcal G}_{85}},\\ \label{d3p}
\overline{{\mathcal G}_2}\;\overline{{\mathcal G}_8}
&=&
\overline{{\mathcal G}_{49}}+\overline{{\mathcal G}_{29}}+\overline{{\mathcal G}_{30}}
+\overline{{\mathcal G}_{37}}+\overline{{\mathcal G}_{39}}
+\overline{{\mathcal G}_{42}}+\overline{{\mathcal G}_{45}},\\ \label{d3q}
\overline{{\mathcal G}_1}\;\overline{{\mathcal G}_9}
&=&
\overline{{\mathcal G}_{23}}+\overline{{\mathcal G}_{24}}+\overline{{\mathcal G}_{25}},\\ \label{d3r}
\overline{{\mathcal G}_2}\;\overline{{\mathcal G}_9}
&=&
\overline{{\mathcal G}_{58}}+\overline{{\mathcal G}_{59}}+\overline{{\mathcal G}_{61}},\\ \label{d3s}
\overline{{\mathcal G}_2}\;\overline{{\mathcal G}_{12}}
&=&
\overline{{\mathcal G}_{59}}+3\overline{{\mathcal G}_{68}}+3\overline{{\mathcal G}_{70}}
+2\overline{{\mathcal G}_{71}}+\overline{{\mathcal G}_{73}},\\ \label{d3t}
\overline{{\mathcal G}_1}\;\overline{{\mathcal G}_{12}}
&=&
\overline{{\mathcal G}_{26}}+\overline{{\mathcal G}_{33}}+\overline{{\mathcal G}_{34}}
+\overline{{\mathcal G}_{35}}+\overline{{\mathcal G}_{36}},\\ \label{d3u}
\overline{{\mathcal G}_1}\;\overline{{\mathcal G}_{14}}
&=&
\overline{{\mathcal G}_{28}}+2\overline{{\mathcal G}_{34}}+\overline{{\mathcal G}_{37}}
+\overline{{\mathcal G}_{40}}+\overline{{\mathcal G}_{41}},\\ \label{d3v}
\overline{{\mathcal G}_2}\;\overline{{\mathcal G}_{14}}
&=&
\overline{{\mathcal G}_{60}}+\overline{{\mathcal G}_{63}}+\overline{{\mathcal G}_{69}}
+\overline{{\mathcal G}_{76}}+\overline{{\mathcal G}_{78}},\\ \label{d3w}
\overline{{\mathcal G}_1}\;\overline{{\mathcal G}_{17}}
&=&
\overline{{\mathcal G}_{31}}+\overline{{\mathcal G}_{44}}+\overline{{\mathcal G}_{46}}
+\overline{{\mathcal G}_{47}},\\ \label{d3x}
\overline{{\mathcal G}_2}\;\overline{{\mathcal G}_{17}}
&=&
\overline{{\mathcal G}_{61}}+\overline{{\mathcal G}_{71}}+2\overline{{\mathcal G}_{73}}
+3\overline{{\mathcal G}_{81}},\\ \label{d3y}
\overline{{\mathcal G}_1}\;\overline{{\mathcal G}_{21}}
&=&
\overline{{\mathcal G}_{42}}+2\overline{{\mathcal G}_{51}}+\overline{{\mathcal G}_{53}}
+\overline{{\mathcal G}_{55}},\\ \label{d3z}
\overline{{\mathcal G}_2}\;\overline{{\mathcal G}_{21}}
&=&
\overline{{\mathcal G}_{65}}+2\overline{{\mathcal G}_{74}}+\overline{{\mathcal G}_{77}}
+\overline{{\mathcal G}_{82}}.
\end{eqnarray}

From these equations one can derive further ones for multiple products, for example:

\begin{eqnarray} \nonumber
\overline{{\mathcal G}_1}\;\overline{{\mathcal G}_{2}}^2
&=&
\overline{{\mathcal G}_{23}}+\overline{{\mathcal G}_{24}}\\ \label{d3aa}
&&+2\left(\overline{{\mathcal G}_{26}}
+\overline{{\mathcal G}_{33}}+\overline{{\mathcal G}_{34}}+\overline{{\mathcal G}_{35}}+\overline{{\mathcal G}_{36}}
+\overline{{\mathcal G}_{31}}+\overline{{\mathcal G}_{44}}+\overline{{\mathcal G}_{47}}\right).
\end{eqnarray}

\section{Results}\label{sec:R}
\subsection{Moments}\label{sec:R_mom}
It turns out that the moments $t_n$ can be written in the following way:
\begin{equation}\label{r1}
t_n=\sum_{\nu\in T_n}\;\overline{\mathcal G}_\nu\; p_\nu(r) \;.
\end{equation}
Here the ${\mathcal G}_\nu,\,\nu\in T_n,$ denote certain
multigraphs of size $n$ and the $p_\nu$ are polynomials of order
$\le n$ in the variable $r=s(s+1)$. It is crucial that these
polynomials depend neither on $N$ nor on the coupling constants
$J_{\mu\nu}$ whereas the terms $\overline{{\mathcal G}_\nu}$
depend only on the coupling constants and only implicitly on $N$
via (\ref{d2}). For the determination of $t_n$ it thus suffices to
enumerate the multigraphs ${\mathcal G}_\nu, \;\nu\in T_n$ and the
corresponding polynomials $p_\nu$. The first moments are
well-known:
\begin{eqnarray}\label{r2a}
t_1&=&0\\ \label{r2b}
t_2&=& \sum_{\mu<\nu}J_{\mu\nu}^2\;\frac{1}{3}r^2= \overline{\;\;\dii}\quad\frac{1}{3}r^2=
\overline{{\mathcal G}_2}\quad\frac{1}{3}r^2
\end{eqnarray}

The third moment $t_3$ has been calculated in \cite{SSL01}. We
reproduce this result using our notation and the multigraphs represented in table \ref{table1}.
\begin{eqnarray}\label{r3}
t_3&=&
-\frac{1}{6} r^2\; \overline{{\mathcal G}_4} + \frac{2}{3} r^3 \;\overline{{\mathcal G}_7}
\;.
\end{eqnarray}

The next moments
$t_4,\,t_5,\,t_6$ have, to our best knowledge, not yet been published,
 although
the polynomials which appear in these expressions are known up
to $8$th order, see \cite{DG74}.

 They are given by the following expressions:

\begin{eqnarray} \nonumber
t_4&=& \frac{1}{15}r^2 (2 + r (-2 + 3 r))\; \overline{{\mathcal
G}_9} +\frac{2}{9}r^3(-1 + 3 r)\; \overline{{\mathcal G}_{12}}
-\frac{2}{9} r^3\;\overline{{\mathcal G}_{14}}\\ \label{r4} &&
+\frac{2}{3} r^4\;\overline{{\mathcal G}_{17}} +\frac{8}{9}
r^4\;\overline{{\mathcal G}_{21}} \;,
\end{eqnarray}

\begin{eqnarray}\nonumber
t_5&=&
-\frac{1}{6} r^2 (1 + 2 (-1 + r) r)\;\overline{{\mathcal G}_{23}}
+\frac{5}{18} r^3 (1 - 2 r)\;\overline{{\mathcal G}_{26}}
\\ \nonumber
&&
+\frac{1}{9} r^3 (3 - 8 r + 12 r^2)\;\overline{{\mathcal G}_{28}}
-\frac{5}{9} r^4 \;\overline{{\mathcal G}_{31}}
+\frac{20}{27} r^4 (-1 + 3 r) \;\overline{{\mathcal G}_{38}}
\\ \label{r5}
&&
-\frac{10}{27} r^4  \;\overline{{\mathcal G}_{42}}
+\frac{20}{9} r^5  \;\overline{{\mathcal G}_{48}}
+\frac{40}{27} r^5  \;\overline{{\mathcal G}_{56}}
\;,
\end{eqnarray}

and

\begin{eqnarray}\nonumber
t_6&=&
\frac{1}{105} r^2 (32 - 87 r + 88 r^2 - 30 r^3 + 15 r^4)\;\overline{{\mathcal G}_{58}}
+\frac{1}{6} r^3 (-3 + 8 r - 8 r^2 + 6 r^3)\;\overline{{\mathcal G}_{59}}
\\ \nonumber
&&
-\frac{2}{3} r^3 (1 - 3 r + 3 r^2)\;\overline{{\mathcal G}_{60}}
+\frac{1}{3} r^4 (2 - 2 r + 3 r^2)\;\overline{{\mathcal G}_{61}}
+\frac{2}{9} r^3 (-2 + 3 r) \;\overline{{\mathcal G}_{62}}
\\ \nonumber
&&
+\frac{1}{3} r^4 (1-2r) \;\overline{{\mathcal G}_{63}}
+\frac{10}{9} r^4 (1-2r)  \;\overline{{\mathcal G}_{64}}
+\frac{2}{9} r^4 (3 - 8 r + 12 r^2) \;\overline{{\mathcal G}_{65}}
\\ \nonumber
&&
+\frac{5}{9} r^4  \;\overline{{\mathcal G}_{66}}
-\frac{20}{9} r^5  \;\overline{{\mathcal G}_{67}}
+\frac{10}{9} r^4 (1 - 3 r + 3 r^2) \;\overline{{\mathcal G}_{68}}
\\ \nonumber
&&
+\frac{5}{9} r^4 (1-2r) \;\overline{{\mathcal G}_{69}}
+\frac{1}{5} r^3 (-1 + 9 r - 22 r^2 + 22 r^3)  \;\overline{{\mathcal G}_{70}}
\\ \nonumber
&&
+\frac{2}{9} r^4 (2 - 10 r + 15 r^2) \;\overline{{\mathcal G}_{71}}
+\frac{2}{9} r^4  \;\overline{{\mathcal G}_{72}}
+\frac{10}{9} r^5 (-1 + 3 r) \;\overline{{\mathcal G}_{73}}
\\ \nonumber
&&
+\frac{2}{9} r^4 (5 - 24 r + 36 r^2) \;\overline{{\mathcal G}_{74}}
-\frac{2}{9} r^4 \;\overline{{\mathcal G}_{75}}
+\frac{2}{9} r^4 (1 - 5 r)  \;\overline{{\mathcal G}_{76}}
\\ \nonumber
&&
+\frac{40}{27} r^5 (-1 + 3 r) \;\overline{{\mathcal G}_{77}}
-\frac{10}{9} r^5 \;\overline{{\mathcal G}_{78}}
+\frac{1}{3} r^4   \;\overline{{\mathcal G}_{79}}
-\frac{20}{27} r^5 \;\overline{{\mathcal G}_{80}}
\\ \nonumber
&&
+\frac{10}{3} r^6 \;\overline{{\mathcal G}_{81}}
+\frac{40}{9} r^6   \;\overline{{\mathcal G}_{82}}
+\frac{80}{27} r^5 (-1 + 3 r)\;\overline{{\mathcal G}_{83}}
\\ \label{r6}
&&
+\frac{2}{9} r^4 \;\overline{{\mathcal G}_{84}}
+\frac{80}{9} r^6   \;\overline{{\mathcal G}_{85}}
+\frac{80}{27} r^6 \;\overline{{\mathcal G}_{86}}
\;.
\end{eqnarray}

\subsection{Free energy}\label{sec:R_f}
It is well-known that the coefficients of the power series for the free energy $F(\beta)$
\begin{equation}\label{rf1}
-\beta F(\beta)=\ln\left( \mbox{Tr } e^{-\beta H}\right)=\sum_{n=0}^{\infty} a_n \beta^n
\end{equation}
can be expressed in terms of the moments $t_n$ and its products.
As indicated in section \ref{sec:D}, a variety of product rules can be used to simplify the resulting expressions.
This simplification, which is sometimes also referred to as the ``cumulant expansion", see \cite{OHZ06},
has the further advantage that it reveals the extensive character of the $a_n$,
see section \ref{sec:R_susc} for a more detailed discussion. The first seven coefficients of the series (\ref{rf1})
read as follows.

\begin{eqnarray} \label{rf2a}
a_0&=&N\,\ln(2s+1),\\ \label{rf2b}
a_1&=&0, \\ \label{rf2c}
a_2&=&\frac{1}{6}r^2\; \overline{{\mathcal G}_2},\\ \label{rf2d}
a_3&=&\frac{1}{36}r^2 \; \overline{{\mathcal G}_4}
-\frac{1}{9}r^3\; \overline{{\mathcal G}_{7}},\\ \label{rf2e}
a_4&=&-\frac{1}{180}r^2(-1 + r + r^2) \; \overline{{\mathcal G}_9}
-\frac{1}{108}r^3\; \overline{{\mathcal G}_{12}}
-\frac{1}{108}r^3\; \overline{{\mathcal G}_{14}}
+\frac{1}{27}r^4\; \overline{{\mathcal G}_{21}},\\ \nonumber
a_5&=&-\frac{1}{2160}r^2(-3 + 6 r + 4 r^2) \; \overline{{\mathcal G}_{23}}
-\frac{1}{432}r^3\; \overline{{\mathcal G}_{26}}\\ \nonumber
&&
+\frac{1}{1080}r^3(-3 + 8 r + 8 r^2)\; \overline{{\mathcal G}_{28}}\\ \label{rf2f}
&&+\frac{1}{162}r^4\; \overline{{\mathcal G}_{38}}
+\frac{1}{324}r^4\; \overline{{\mathcal G}_{42}}
-\frac{1}{81}r^5\; \overline{{\mathcal G}_{56}},\\ \nonumber
a_6&=&\frac{1}{453600}r^2(192 - 522 r - 67 r^2 + 240 r^3 + 160 r^4) \; \overline{{\mathcal G}_{58}}\\ \nonumber
&&
+\frac{1}{12960}r^3(-9 + 12 r + 8 r^2) \; \overline{{\mathcal G}_{59}}
+\frac{1}{1080}r^3(-1 + 3 r + 2 r^2)\; \overline{{\mathcal G}_{60}}\\ \nonumber
&&
+\frac{1}{6480}r^3(-4 + r)\; \overline{{\mathcal G}_{62}}
+\frac{1}{6480}r^4(3 + 4 r)\; \overline{{\mathcal G}_{63}}
+\frac{1}{648}r^4\; \overline{{\mathcal G}_{64}}\\ \nonumber
&&
-\frac{1}{3240}r^4(-3 + 8 r + 8 r^2)\; \overline{{\mathcal G}_{65}}
+\frac{1}{648}r^4\; \overline{{\mathcal G}_{68}}\\ \nonumber
&&
+\frac{1}{1296}r^4\; \overline{{\mathcal G}_{69}}
-\frac{1}{32400}r^3(9 - 81 r + 48 r^2 +152 r^3)\; \overline{{\mathcal G}_{70}}\\ \nonumber
&&
+\frac{1}{1620}r^4\; \overline{{\mathcal G}_{71}}
+\frac{1}{3240}r^4\; \overline{{\mathcal G}_{72}}\\ \nonumber
&&
-\frac{1}{3240}r^4(-5 + 24 r + 24 r^2) \; \overline{{\mathcal G}_{74}}
-\frac{1}{3240}r^4 \; \overline{{\mathcal G}_{75}}\\ \nonumber
&&
+\frac{1}{3240}r^4 \; \overline{{\mathcal G}_{76}}
-\frac{1}{486}r^5 \; \overline{{\mathcal G}_{77}}
+\frac{1}{2160}r^4 \; \overline{{\mathcal G}_{79}}
-\frac{1}{927}r^5 \; \overline{{\mathcal G}_{80}}\\ \label{rf2g}
&&
-\frac{1}{243}r^5 \; \overline{{\mathcal G}_{83}}
+\frac{1}{3240}r^4 \; \overline{{\mathcal G}_{84}}
+\frac{1}{243}r^6 \; \overline{{\mathcal G}_{86}}
\;.
\end{eqnarray}

\subsection{Magnetic moments and susceptibility}\label{sec:R_susc}
To obtain the magnetic moments $\mu_n$ we can use a special method which is available
if one knows the moments $t_n$ for all values of the coupling constants $J_{\mu\nu}$.
We replace $H$ by the one parameter family
of Hamiltonians $H_\alpha\equiv H+\frac{\alpha}{2}\left({\bf S}^2-N r \right)$.
Equivalently we can substitute $J_{\mu\nu}\mapsto J_{\mu\nu}+\alpha$ for all
coupling constants. The magnetic moments then result from differentiating
$\mbox{Tr}(H_\alpha^{n+1})$ w.~r.~t.~$\alpha$ and finally setting $\alpha=0$:

\begin{eqnarray}\label{r7a}
\left.\frac{\partial}{\partial\alpha}\;\mbox{Tr}\left(H_\alpha^{n+1}\right)\right|_{\alpha=0}&=&
\frac{n+1}{2}\mbox{Tr}\left(H_0^n({\bf S}^2-N r)\right)\\ \label{r7b}
&=&\frac{(n+1)(2s+1)^N}{2}\left( 3\mu_n - N r t_n\right)
\;.
\end{eqnarray}

We can calculate the left hand side of (\ref{r7a}) if we insert the results for the moments and consider ``derivatives"
${\mathcal G}'$ of multigraphs
defined in the following way. Let ${\mathcal G}^{(ij)}$ denote the multigraph ${\mathcal G}$ but with one bond
removed, ${\mathcal N}(i,j)\mapsto {\mathcal N}(i,j)-1$. If ${\mathcal N}(i,j)=0$ then we set ${\mathcal G}^{(ij)}=0$.
Further let $G({\mathcal G})$ and $G({\mathcal G}^{(ij)})$ denote the respective symmetry groups. Then we define
\begin{equation}\label{r8}
{\mathcal G}' = \sum_{i<j}{\mathcal N}(i,j)\;{\mathcal G}^{(ij)}
\;\frac{|G({\mathcal G})|}{|G({\mathcal G}^{(ij)})|}
\;.
\end{equation}
One has, so to speak, to break each bond of the multigraph and to sum over all results. Further, one has to introduce
factors which compensate for the possible change of symmetries. For example,
$\,\tiiii \,'= 6\; \ti\;+\;\tii\;$. It is obvious that the evaluation of ${\mathcal G}'$ just yields
$\left.\frac{\partial}{\partial\alpha}\;\overline{\mathcal G}\right|_{\alpha=0}$.
Then it is a straightforward task to calculate the magnetic moments $\mu_0,\ldots,\mu_5$ by using the
above results for the $t_n$. We will not, however, display these results and pass to the $c_n$. \\

The coefficients of the high temperature expansion of
$\chi=\beta \frac{\mbox{\scriptsize Tr}(\mathbf{S}^{(3)2}\exp(-\beta H))}{\mbox{\scriptsize Tr}(\exp(-\beta H))}$
can be expressed through the $\mu_n$ and the $t_n$ which occur as coefficients of the series in the numerator or in the
denominator, respectively. The first $6$ coefficients are given by:
\begin{eqnarray}
\chi&=&\sum_{n=1}^\infty c_n\;\beta^n\\ \nonumber
&=&\mu_0\;\beta-\mu_1\;\beta^2 + \frac{1}{2}(\mu_2-\mu_0\,t_2)\beta^3+ \frac{1}{6}(t_3\mu_0+3 t_2 \mu_1-\mu_3)\beta^4\\
\nonumber
&+& \frac{1}{24}(6 t_2^2 \mu_0-t_4\mu_0 - 4 t_3 \mu_1-6 t_2\mu_2+\mu_4)\beta^5\\ \label{r9}
&+&\frac{1}{120}( t_5 \mu_0-30 t_2^2 \mu_1+5 t_4\mu_1+10 t_3 \mu_3+10 t_2(-2 t_3\mu_0+\mu_3)-\mu_5)\beta^6\\ \nonumber
&+&\ldots.
\end{eqnarray}

Inserting the known values for the $t_n$ and the $\mu_n$ yields the desired results for the $c_n$.
Similarly as in section \ref{sec:R_f}, a variety of product rules can be used to simplify the resulting
expressions revealing the extensive character of the $c_n$.
By this we mean the following.
If the spin system under consideration would have a periodic lattice structure of, say, $K$ lattice units
with periodic boundary conditions, it follows immediately that the evaluation of a single multigraph
$\overline{{\mathcal G}}$ linearly scales with $K$, and hence with $N$, as long as ${\mathcal G}$ is connected.
For unconnected ${\mathcal G}$ the evaluation scales with $K^c$ where $c$ is the number of connected
components of ${\mathcal G}$.
Obviously, products of
evaluations of connected multigraphs $\overline{{\mathcal G}_\nu}\;\overline{{\mathcal G}_\mu}$ would scale with $K^2$.
It turns out that the elimination of these and higher products in the expression for the $c_n$
by means of the rules (\ref{d3a})-(\ref{d3r}) also
eliminates the evaluation terms of unconnected multigraphs. This has to be expected on physical grounds,
since the total susceptibility of a spin lattice should be an extensive quantity, i.~e.~linearly scale with $K$.
But it is an additional consistency test of our results that the non-extensive contributions to the $c_n$ actually cancel.\\

We will now represent the results for the susceptibility's HTE up to sixth order in the inverse temperature $\beta$.
$c_1$ and $c_2$ are well-known, $c_3$ has already been published in \cite{SSL01},
but $c_4,c_5$ and $c_6$ seem to be new,
although
the polynomials which appear in these expressions are known up
to $8$th order, see \cite{DG74}.

\begin{eqnarray}\label{r10a}
c_1&=&\frac{Nr}{3},\\ \label{r10b}
c_2&=& -\frac{2}{9} r^2 \,\overline{{\mathcal G}_1},\\ \label{r10c}
c_3&=& -\frac{1}{18} r^2 \,\overline{{\mathcal G}_2}+\frac{2}{27} r^3 \,\overline{{\mathcal G}_3},\\ \label{r10d}
c_4
&=&
\frac{2}{135} r^2 (-1 + r + r^2)\,\overline{{\mathcal G}_4}
+\frac{1}{54} r^3 \,\overline{{\mathcal G}_5}+\frac{1}{27} r^3 \,\overline{{\mathcal G}_7}
-\frac{2}{81} r^4 \, \overline{{\mathcal G}_8},\\ \nonumber
c_5
&=&
\frac{1}{648} r^2 (-3 + 6 r + 4 r^2)\,\overline{{\mathcal G}_9}
-\frac{2}{405} r^3 (-1 + r + r^2)\,\overline{{\mathcal G}_{10}}\\ \nonumber
&&+\frac{1}{108} r^3 \,\overline{{\mathcal G}_{12}}
-\frac{1}{243} r^4\,\overline{{\mathcal G}_{13}}
-\frac{1}{540} r^3(-3 + 8 r + 8 r^2)\,\overline{{\mathcal G}_{14}}\\ \label{r10e}
&&
-\frac{1}{486} r^4\,\overline{{\mathcal G}_{15}}
-\frac{1}{162} r^4\,\overline{{\mathcal G}_{16}}
-\frac{2}{243} r^4\,\overline{{\mathcal G}_{19}}
-\frac{4}{243} r^4\,\overline{{\mathcal G}_{21}}
+\frac{2}{243} r^5\,\overline{{\mathcal G}_{22}},\\ \nonumber
c_6
&=&
-\frac{1}{113400} r^2 (192 - 522 r - 67 r^2 + 240 r^3 + 160 r^4)\,\overline{{\mathcal G}_{23}}  \\ \nonumber
&&
-\frac{1}{1944} r^3 (-3 + 6 r + 4 r^2)\,\overline{{\mathcal G}_{24}}
-\frac{1}{972} r^3 (-3 + 3 r + 2 r^2)\,\overline{{\mathcal G}_{26}} \\  \nonumber
&&
-\frac{1}{972} r^4 \,\overline{{\mathcal G}_{27}}
-\frac{1}{2430} r^3 (-6 + 21 r + 16 r^2)\,\overline{{\mathcal G}_{28}}\\ \nonumber
&&
+\frac{1}{4860} r^4 (-3+ 8 r + 8 r^2)\,\overline{{\mathcal G}_{29}}
+\frac{2}{1215} r^4 (-1 + r + r^2) \,\overline{{\mathcal G}_{30}}\\ \nonumber
&&
-\frac{1}{324} r^4\,\overline{{\mathcal G}_{33}}
+\frac{1}{24300} r^3 (9 - 126 r - 12 r^2 + 152 r^3)\,\overline{{\mathcal G}_{34}}
-\frac{1}{648} r^4 \,\overline{{\mathcal G}_{35}}\\ \nonumber
&&
+\frac{1}{1620} r^4 (-3 + 8 r + 8 r^2)\,\overline{{\mathcal G}_{37}}
-\frac{5}{972} r^4 \,\overline{{\mathcal G}_{38}}
+\frac{1}{729} r^5 \,\overline{{\mathcal G}_{39}}\\ \nonumber
&&
+\frac{1}{2430} r^4 (-7 + 12 r + 12 r^2) \,\overline{{\mathcal G}_{42}}
+\frac{1}{1458} r^5 \,\overline{{\mathcal G}_{43}}
-\frac{1}{648} r^4 \,\overline{{\mathcal G}_{44}}\\ \nonumber
&&
+\frac{1}{486} r^5 \,\overline{{\mathcal G}_{45}}
+\frac{2}{729} r^5 \,\overline{{\mathcal G}_{50}}
+\frac{1}{1620} r^4 (-1 + 16 r + 16 r^2)\,\overline{{\mathcal G}_{51}}\\ \label{r10f}
&&
+\frac{2}{729} r^5 \,\overline{{\mathcal G}_{52}}
+\frac{2}{729} r^5 \,\overline{{\mathcal G}_{53}}
+\frac{5}{729} r^5 \,\overline{{\mathcal G}_{56}}
-\frac{2}{729} r^6 \,\overline{{\mathcal G}_{57}}
\;.
\end{eqnarray}


\newpage
\section*{References}
\begin{thebibliography}{10}

\bibitem{DG74} G.S.~Rushbrooke, G.A. Baker, and P.J. Wood, in
{\em Phase Transitions and Critical Phenomena}, Vol. 3, p. 245; eds.
C.~Domb and M.S.~Green, Academic Press, London, 1974.


\bibitem{OHZ06}
J~Oitmaa, CJ~Hamer, and WH~Zheng.
\newblock {\em {Series Expansion Methods}}.
\newblock {Cambridge University Press}, {2006}.

\bibitem{SSL01}
HJ~Schmidt, J~Schnack, and M~Luban.
\newblock {Heisenberg exchange parameters of molecular magnets from the
  high-temperature susceptibility expansion}.
\newblock {\em {Phys.~Rev.~B}}, {64}({22}), {DEC 1} {2001}.

\bibitem{TWB10}
CA~Thuesen, H~Weihe, J~Bendix, S~Piligkos, and O~Monsted.
\newblock {Computationally inexpensive interpretation of magnetic data for
  finite spin clusters}.
\newblock {\em {Dalton Trans.}}, {39}({20}):{4882--4885}, {2010}.

\bibitem{CGSM08}
PJ~Cregg, JL~Garcia-Palacios, P~Svedlindh, and K~Murphy.
\newblock {Low-field susceptibility of classical Heisenberg chains with
  arbitrary and different nearest-neighbour exchange}.
\newblock {\em {J.~Phys.-Cond.~Mat.}}, {20}({20}), {MAY 21}
  {2008}.

\bibitem{FHW04}
N~Fukushima, A~Honecker, S~Wessel, and W~Brenig.
\newblock {Thermodynamic properties of ferromagnetic mixed-spin chain systems}.
\newblock {\em {Phys.~Rev.~B}}, {69}({17}), {MAY} {2004}.

\bibitem{OZ04a}
J~Oitmaa and WH~Zheng.
\newblock {Phase diagram of the bcc S=1/2 Heisenberg antiferromagnet with first
  and second neighbor exchange}.
\newblock {\em {Phys.~Rev.~B}}, {69}({6}), {FEB} {2004}.

\bibitem{OZ04b}
J~Oitmaa and WH~Zheng.
\newblock {Curie and Neel temperatures of quantum magnets}.
\newblock {\em {J.~Phys.-Cond.~Mat.}}, {16}({47}):{8653--8660},
  {DEC 1} {2004}.

\bibitem{LKMW03}
M~Luban, P~Kogerler, LL~Miller, and REP Winpenny.
\newblock {Heisenberg model of a \{Cr-8\}-cubane magnetic molecule}.
\newblock {\em {J.~Appl.~Phys.}}, {93}({10, Part 2}):{7083--7085},
  {MAY 15} {2003}.

\bibitem{ST02}
M~Shiroishi and M~Takahashi.
\newblock {Integral equation generates high-temperature expansion of the
  Heisenberg chain}.
\newblock {\em {Phys.~Rev.~Lett.}}, {89}({11}), {SEP 9} {2002}.

\bibitem{HL01}
A~Honecker and A~Lauchli.
\newblock {Frustrated trimer chain model and Cu3Cl6(H2O)(2)center dot
  2H(8)C(4)SO(2) in a magnetic field}.
\newblock {\em {Phys.~Rev.~B}}, {63}({17}), {MAY 1} {2001}.

\bibitem{BEU00}
A~Buhler, N~Elstner, and GS~Uhrig.
\newblock {High temperature expansion for frustrated and unfrustrated S=1/2
  spin chains}.
\newblock {\em {Eur.~Phys.~J.~B}}, {16}({3}):{475--486}, {AUG}
  {2000}.

\bibitem{ZHO99}
WH~Zheng, CJ~Hamer, and J~Oitmaa.
\newblock {Series expansions for a Heisenberg antiferromagnetic model for
  SrCu2(BO3)(2)}.
\newblock {\em {Phys.~Rev.~B}}, {60}({9}):{6608--6616}, {SEP 1} {1999}.

\bibitem{ES98a}
N~Elstner and RRP Singh.
\newblock {Strong-coupling expansions at finite temperatures: Application to
  quantum disordered and quantum critical phases}.
\newblock {\em {Phys.~Rev.~B}}, {57}({13}):{7740--7748}, {APR 1} {1998}.

\bibitem{ES98b}
N~Elstner and RRP Singh.
\newblock {Field-dependent thermodynamics and quantum critical phenomena in the
  dimerized spin system Cu-2(C5H12N2)(2)Cl-4}.
\newblock {\em {Phys.~Rev.~B}}, {58}({17}):{11484--11487}, {NOV 1} {1998}.

\bibitem{OB96}
J~Oitmaa and E~Bornilla.
\newblock {High-temperature-series study of the spin-1/2 Heisenberg
  ferromagnet}.
\newblock {\em {Phys.~Rev.~B}}, {53}({21}):{14228--14235}, {JUN 1} {1996}.

\bibitem{KBJ96}
J~Karwowski, D~BielinskaWaz, and J~Jurkowski.
\newblock {Eigenvalues of model Hamiltonian matrices from spectral density
  distribution moments: The Heisenberg spin Hamiltonian}.
\newblock {\em {Int.~J.~Quant.~Chem.}},
  {60}({1}):{185--193}, {OCT 5} {1996}.

\bibitem{Jetal00}
DC~Johnston, RK~Kremer, M~Troyer, X~Wang, A~Klumper, SL~Bud'ko,
AF~Panchula,
  and PC~Canfield.
\newblock {Thermodynamics of spin S=1/2 antiferromagnetic uniform and
  alternating-exchange Heisenberg chains}.
\newblock {\em {Phys.~Rev.~B}}, {61}({14}):{9558--9606}, {APR 1} {2000}.

\bibitem{Retal03}
H~Rosner, RRP Singh, WH~Zheng, J~Oitmaa, and WE~Pickett.
\newblock {High-temperature expansions for the J(1)-J(2) Heisenberg models:
  Applications to ab initio calculated models for Li2VOSiO4 and Li2VOGeO4}.
\newblock {\em {Phys.~Rev.~B}}, {67}({1}), {JAN 1} {2003}.

\bibitem{MBP03}
G~Misguich, B~Bernu, and L~Pierre.
\newblock {Determination of the exchange energies in Li2VOSiO4 from a
  high-temperature series analysis of the square-lattice J(1)-J(2) Heisenberg
  model}.
\newblock {\em {Phys.~Rev.~B}}, {68}({11}), {SEP 15} {2003}.

\end{thebibliography}
\end{document}